\documentclass{jfm}
\usepackage{hyperref}
\hypersetup{
	colorlinks = true,
	allcolors = {blue}
}
\usepackage{textcomp}
\usepackage{natbib}
\usepackage{comment}
\usepackage{amsfonts}
\usepackage{amssymb}
\usepackage{amsmath}
\usepackage{epsfig}
\usepackage{calrsfs}
\usepackage{fancybox}
\usepackage{pstricks}
\usepackage{float}

\usepackage{graphicx}

\usepackage{tikz}

\usepackage{epstopdf}

\usepackage{amsmath} 
\usepackage{color} 
\usepackage{xcolor}

\title{Sliding, vibrating and swinging droplets on an oscillating fibre}
\shorttitle{Sliding, vibrating and swinging droplets on an oscillating fibre}

\author{Stéphane Poulain \corresp{\email{stephapo@math.uio.no}} \and Andreas Carlson \corresp{\email{acarlson@math.uio.no}}}
\shortauthor{S. Poulain and A. Carlson}
\affiliation{ 
	Department of Mathematics, Mechanics Division, University of Oslo, N-0851 Oslo, Norway
}

\begin{document}
\maketitle

\begin{abstract}
We study experimentally the dynamics of a water droplet on a tilted and vertically oscillating rigid fibre. As we vary the frequency and amplitude of the oscillations the droplet transitions between different modes: harmonic pumping, subharmonic pumping, a combination of rocking and pumping modes, and a combination of pumping and swinging modes.
We characterize these responses and report how they affect the droplet's sliding speed along the fibre.
The droplet swinging mode is explained by a minimal model making an analogy between the droplet and a forced elastic pendulum.
\end{abstract}


\section{Introduction}
\label{sec:Intro}

The interactions between liquid drops and fibres is ubiquitous in a wide range of situations including liquid aerosol filtering \citep{Agranovski1998,Zhang2015}, coating processes \citep{Quere1999,Chan2021}, %where controlling a droplet's speed modifies the thickness of the deposited film,
digital microfluidics \citep{Gilet2009,Gilet2010}
and fog harvesting \citep{Klemm2012,Labbe2019}.
The latter has also motivated research of droplets interacting with biological systems \citep{Malik2014} such as threads of spider silk \citep{Zheng2010,Ju2014} and plants with fibre-like features such as sequoia needles, cactus spines, grass blades and moss leaves \citep{Limm2009,Ju2012,Roth2012,Pan2016} that are able to efficiently capture and transport water droplets.
In most of these examples, the fibre is generally not still but subject to motion due to external perturbations such as wind.

Drops can move on horizontal fibres due to spatial gradients in various properties: shape, most notably with conical fibres \citep{Lorenceau2004,McCarthy2019,Chan2020}, wetting \citep{Zheng2010,Ju2014}, temperature \citep{Yarin2002} or elasticity \citep{Duprat2012}.
Droplets on non-horizontal fibres slide when the gravitational force overcomes contact angle hysteresis. %\citep{Huang2009,Darbois2015}.
\citet{Gilet2010} studied this in details for the case when droplets perfectly wet fibres while \citet{Christianto2022} recently investigated numerically the effect of a finite contact angle.
In addition to these passive mechanisms, external perturbations in the form of standing waves \citep{Bick2015} or wind \citep{Dawar2006,Dawar2008,Sahu2013,Bintein2019} also lead to directional transport of droplets on fibres.
There has been anecdotal reports pointing to vibrations triggering  droplet motion on fibres \citep{Dawar2006,Dawar2008,Zhang2018}, yet quantitative data to describe this effect are lacking.

One way to induce reproducible vibrations of droplets is by inducing  rigid-body oscillations of the substrate.
So far studies of this phenomenon have only focused on flat, planar surfaces, where two types of experimental setups have been employed: droplets on a slanted flat substrate submitted to vertical oscillations, and droplets on a horizontal flat substrate submitted to slanted oscillations.
In both cases a directional motion of the droplet takes place for high enough amplitude of oscillations.
Recent reviews of both experimental and numerical results regarding the rich dynamics of these systems are given by \citet{Bradshaw2018,Deegan2020,Costalonga2020}.
In short, a droplet in such a situation experiences a modulation of its contact area through a \emph{pumping} mode of vibrations that periodically stretches and flattens it, while also experiencing \emph{rocking} lateral vibrations. The combination of both rocking and pumping responses, and in particular their phase difference \citep{Noblin2009}, triggers directional motion.
On slanted substrates, a pumping mode alone can trigger motion if the periodic evolution of the wetted area unpins the droplet.

Quantitatively, for a given frequency of vibrations $f$, the amplitude of vibrations $A$ needs to be larger than a threshold $A_{\rm th} \geq 0$ to trigger motion: for $A>A_{\rm th}$, droplets  have a non-zero mean velocity $\langle U \rangle $, defined as the velocity of the center of mass along the fibre averaged over one period of oscillations. $\langle U \rangle$ is typically in the direction of the oscillations or in that of gravity, leading to a sliding droplet with $\langle U \rangle>0$. A less intuitive regime of climbing droplets with $\langle U \rangle<0$ also exists \citep{Brunet2007,Sartori2019,Costalonga2020}.
In the most common case of sliding drops, \citet{Costalonga2020} proposed the following empirical relationship:
\begin{align}
	\langle U \rangle-U_0 = s  \left(A-A_{\rm th}\right)^\chi, \quad A>A_{\rm th}.
	\label{eq:dropmotion}
\end{align}
 We have modified this relation to account for $U_0$, the speed that the droplet has without oscillations: $U_0$ need not vanish on tilted substrates. 
The exponent $\chi$ and the mobility coefficient $s$ quantify the nature of the relationship between the speed and the amplitude.
Their values along with that of $A_{\rm th}$ characterize the response of a droplet on a substrate submitted to vibrations.
These parameters depend on the liquid properties (surface tension coefficient, density, viscosity), the size of the droplet, the wetting properties of the substrate, the frequency $f$ of the oscillations, and the angles of both the substrate and oscillations with respect to the horizontal direction.
\citet{Costalonga2020} highlight that numerical works are typically consistent with  $1\leq \chi \leq 2$  while experimental observations usually suggest $\chi \simeq 1$, yielding a linear relationship between the forcing amplitude and the speed.
Yet values of $\chi$ larger than 1  can be observed, and a regime of decreasing speed upon increase of forcing has also been reported by \citet{Sartori2019}.
Overall the dependence of the coefficients involved in \eqref{eq:dropmotion} are not well understood, even though the experiments by \citet{Costalonga2020} shed some lights on the influence of many of the parameters involved for horizontal flat surfaces with low contact angle hysteresis, namely that of the frequency, viscosity, droplet volume, and angle of vibrations.

In this article we experimentally probe the effect of substrate vibrations in another geometry relevant to a range of applications discussed above: a water droplet on a tilted fibre submitted to vertical oscillations. We report the sliding speed as a function of the amplitude and frequency of the forcing.
Further, we describe transitions between different regimes of sliding that differ from prior observations on flat substrates.

\section{Experimental setup}
\label{subsec:oscillationfreq}

\begin{figure}
	 \centering
	\includegraphics[width=0.8\linewidth]{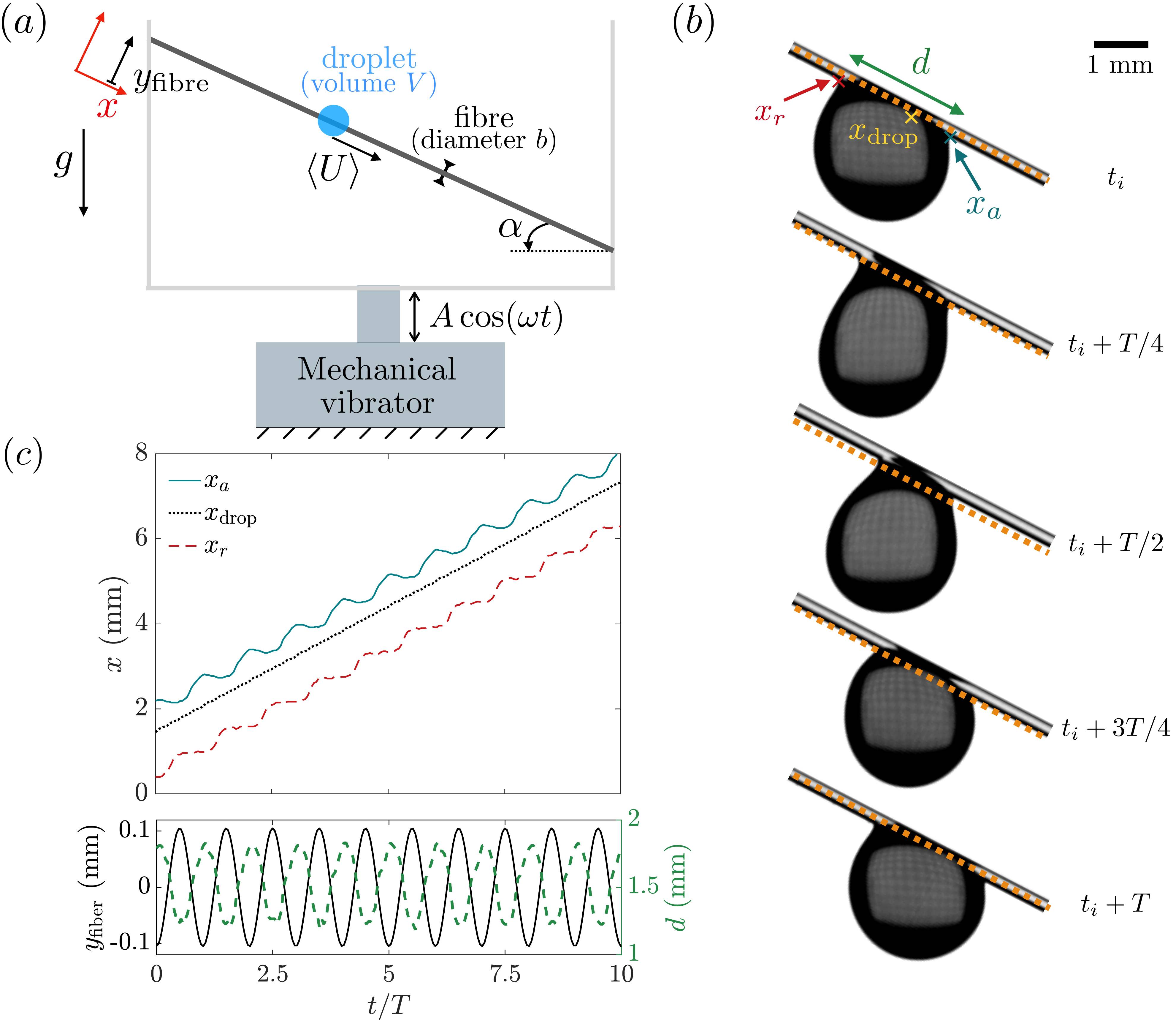}

	\caption{$(a)$ 
	 Schematic of the experimental setup.
	 A well-taut nylon fibre of diameter $b$ is attached to a structure oscillating vertically with amplitude $A$ and angular frequency $\omega=2\pi f$.
	 The fibre makes an angle $\alpha$ with respect to the horizontal direction.
	 As it oscillates and its position evolves as $y_{\rm fibre}=A\cos(\alpha)\cos(\omega t)$, a water droplet of volume $V$ slides down at speed $\langle U \rangle$.
	$(b)$ Images of a droplet over one period $T=1/f$. Here, $V=4~\mu$L, $b=200~\mu$m, $\alpha=27.5^\circ$, $A=0.10$ mm and $f=60$ Hz.  The resulting acceleration from the oscillations is $A\omega^2=14.2$ m.s$^{-2}$ so that  $\Gamma=A\omega^2/g=1.45$. 
	The dotted line represents the minimal value of $y_{\rm fibre}$ and highlights the fibre's motion.
	The position of the center of mass of the droplet projected on the fibre is $x_{\rm drop}$, that of the advancing contact line is $x_a$ and that of the  receding contact line is $x_r$.
	$(c)$ (Top) Time evolution of the position of a droplet with the same conditions as in $(b)$. The droplet moves at near-constant speed \mbox{$\langle U \rangle=\langle {\rm d}x_{\rm drop}/{\rm d}t \rangle = 22.6$ mm.s$^{-1}$}. (Bottom) Corresponding time evolution of the position of the fibre $y_{\rm fibre}$ (solid line) and of the basal diameter $d=x_a-x_r$ (dashed line).
	\label{fig:setup}}
\end{figure}

Our experimental setup is sketched in figure \ref{fig:setup}($a$).
A well-taut nylon fibre (fishing line, Abu Garcia abulon top) of diameter $b$ and making an angle $\alpha$ with the horizontal is connected to a mechanical vibrator (PASCO SF-9324). A periodic sinusoidal signal is generated  (NI myDAQ), amplified (QSC RMX850a) and fed into the vibrator so that the fibre oscillates vertically as $y_{\rm fibre}(t)=A\cos(\alpha)\cos(\omega t)$ where $A$ is the amplitude and $\omega=2\pi f$  the angular frequency of oscillations.
The maximum acceleration of the structure resulting from these oscillations is $A\omega^2$, we normalize it using the gravitational acceleration  $g=9.81$ m.s$^{-2}$ and  let $\Gamma=A\omega^2/g$.
We deposit a droplet of volume $V$, and equivalent spherical radius $r=(3V/4\pi)^{1/3}$, on the oscillating fibre with a micropipette and record its motion over approximately 40 mm along the fibre as it slides downward  using a high speed camera (Photron FASTCAM Mini, frame rate ranging from a few hundreds and up to 5000 fps, typical resolution of 20 $\mu$m.pixel$^{-1}$ using a standard macro lens).
We use deionized water with a small amount of black die (nigrosin) to facilitate visualization; the relevant physical properties of water are its density $\rho=1.0 \times 10^3 $ kg.m$^{-3}$, surface tension coefficient $\sigma=70$ mN.m$^{-1}$ and dynamic viscosity $\mu=1$ mPa.s.
Experiments are performed in air at room temperature (22 $^\circ$C), evaporation is negligible within the timescales involved.
Representative images and measurements are shown in figure \ref{fig:setup}$(b,c)$.

The static contact angle between a water droplet and a nylon fibre is $\theta=65^\circ\pm7^\circ$.
To measure this angle we cut small pieces of the fibre, covered them with a glass slide, and put them in an oven just above the melting point on the material. This result in  glass slide with a uniform and flat layer of nylon. Once cooled to room temperature, we deposited 2 $\mu$L droplets on this substrate. The contact angle was extracted by measuring the angle between the substrate and a high-order polynomial fit of the droplet's boundaries.

The behavior of droplets on oscillating substrates depends strongly on the frequency $f$ of the forcing oscillations, and in particular on the ratio $f/f_{\rm pump}$ with $f_{\rm pump}$ the natural frequency of the first pumping mode of the droplet \citep{Costalonga2020}.
The natural frequencies of vibrations of inviscid droplets scale as $(\sigma/\rho V)^{1/2}$ and the full analytical expression for freely-suspended spherical drops is well-known \citep{Lamb1924}.
Sessile droplets on flat and horizontal surfaces exhibit different responses due to the change of geometry: this has been  studied in details by \citet{Bostwick2015,Chang2015}.
We also expect drops on fibres to exhibit different eigenmodes and natural frequencies that are function of the fibre diameter $b$.
Here, we measure the first natural frequency $f_{\rm pump}$ of pumping oscillations by subjecting a droplet with $V=4~\mu$L ($r\simeq 1$ mm) deposited on a horizontal fibre ($\alpha=0^\circ$) with diameter $b=200~\mu$m to a step vertical acceleration and observing its response: we find $f_{\rm pump}=57 \pm 1$ Hz.

To quantify the expected effect of viscosity in the droplet's behavior, an important parameter is the ratio between the thickness of the Stokes' boundary layer $\delta=(2\mu/\rho\omega)^{1/2}$ and the characteristic size of the droplet, taken here as its  equivalent spherical radius $r=(3V/4\pi)^{1/3}$.
Even for the smallest frequency we have used the Stokes layer is much thinner than the droplet itself: $\delta \approx 0.15~{\rm mm} \ll r \approx 1$ mm for a $V=4~\mu$L water droplet on a fibre oscillating at $f=15$ Hz. This is equivalent to a large Reynolds number $\mathrm{Re}=\rho (r\omega)r/\mu=2(r/\delta)^2\gtrsim 100$.
This suggests that viscous effects are localized in a thin boundary layer and that the flow in most of the droplet is inertial.
Another measure of the importance of viscosity is the Ohnesorge number ${\rm Oh}=\mu(\rho r \sigma)^{-1/2}$, the inverse squared Reynolds number based on the capillary speed $\sigma/\mu$, which compares viscous effects to both inertial and capillary ones; here ${\rm Oh}\approx 4\times 10^{-3} \ll 1$.
The Weber number comparing inertial to capillary effects is $\mathrm{We}=\rho(r\omega)^2r/\sigma$ and range from $0.1$ to $10$ upon varying the frequency from 15 to 135 Hz, suggesting a competition between inertia and capillarity.
We consider droplets that are larger than the fibre they are deposited on, $r>b$, but smaller than the capillary length $l_c=(\sigma/\rho g)^{1/2} \approx 3$ mm so that the Bond number quantifying the ratio of capillary to gravitational effects is ${\rm Bo}=(r/l_c)^2 \approx 0.1$.
While smaller than 1, this Bond number is large enough for gravity to significantly modify the equilibrium shape \citep{Gupta2021} and the droplets we study hang below the fibre (figure \ref{fig:setup}$b$).

\section{Sliding speed}
\label{sec:slidingdrops}

\begin{figure}
	 \centering
	\begin{tikzpicture}
    	\draw (0, 0) node[inner sep=0] (fig) {\includegraphics[width=0.47\textwidth]{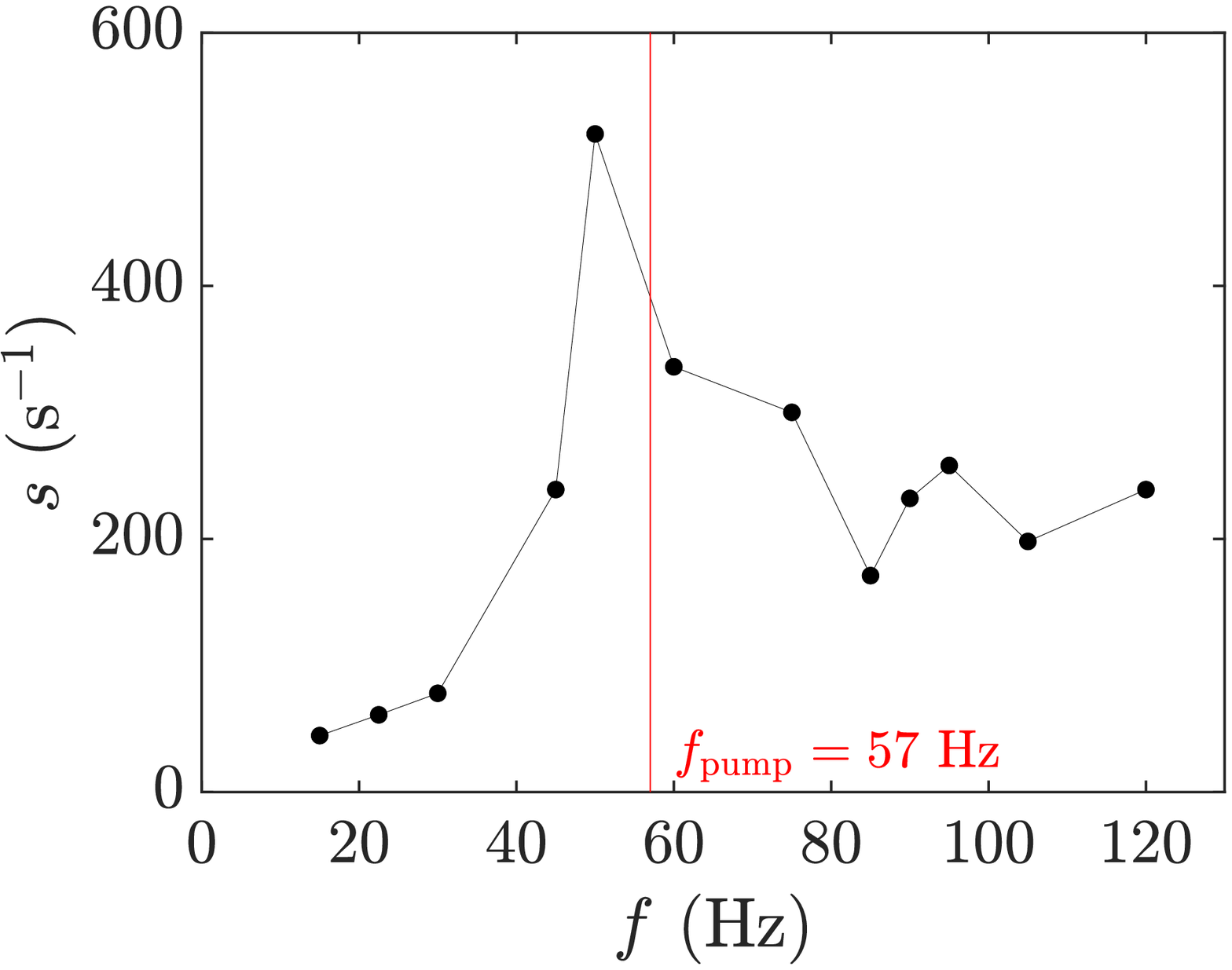}};
    	\node[] at (fig.north west){$(a)$};
  	\end{tikzpicture}
  	\begin{tikzpicture}
    	\draw (0, 0) node[inner sep=0] (fig) {\includegraphics[width=0.47\textwidth]{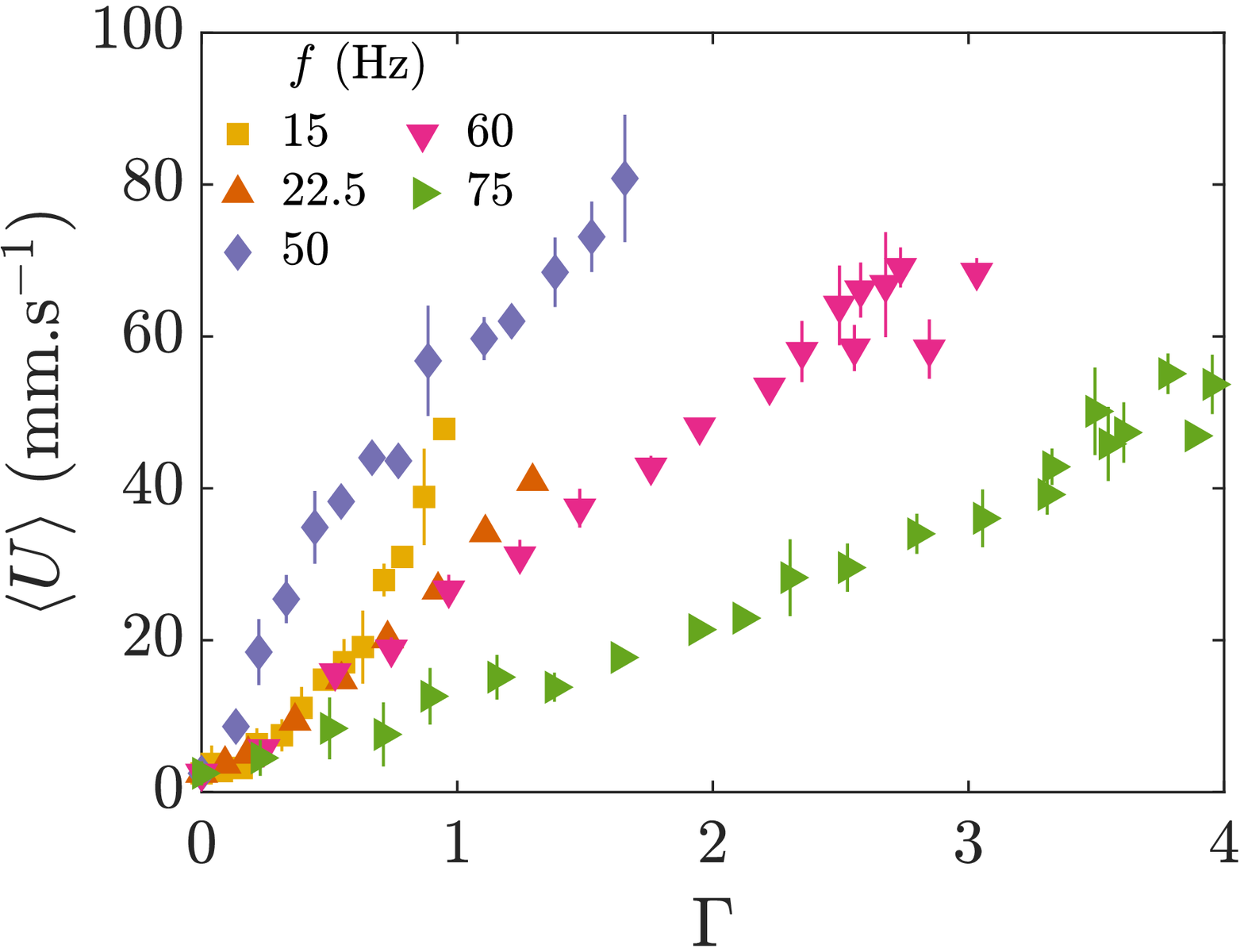}};
    	\node[] at (fig.north west){$(b)$};
  	\end{tikzpicture}

  	\begin{tikzpicture}
    	\draw (0, 0) node[inner sep=0] (fig) {\includegraphics[width=0.47\textwidth]{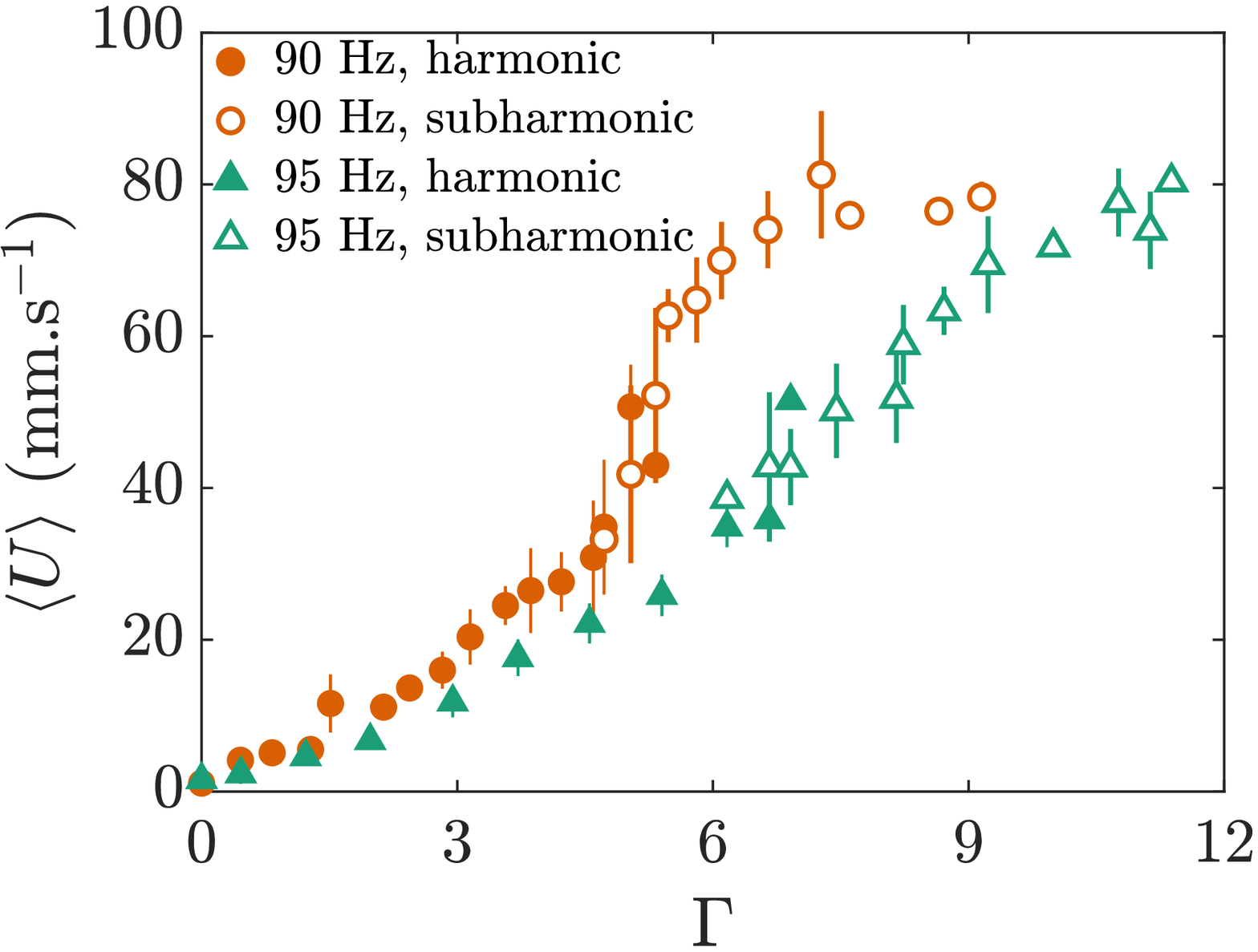}};
    	\node[] at (fig.north west){$(c)$};
  	\end{tikzpicture}
  	\begin{tikzpicture}
    	\draw (0, 0) node[inner sep=0] (fig) {\includegraphics[width=0.47\textwidth]{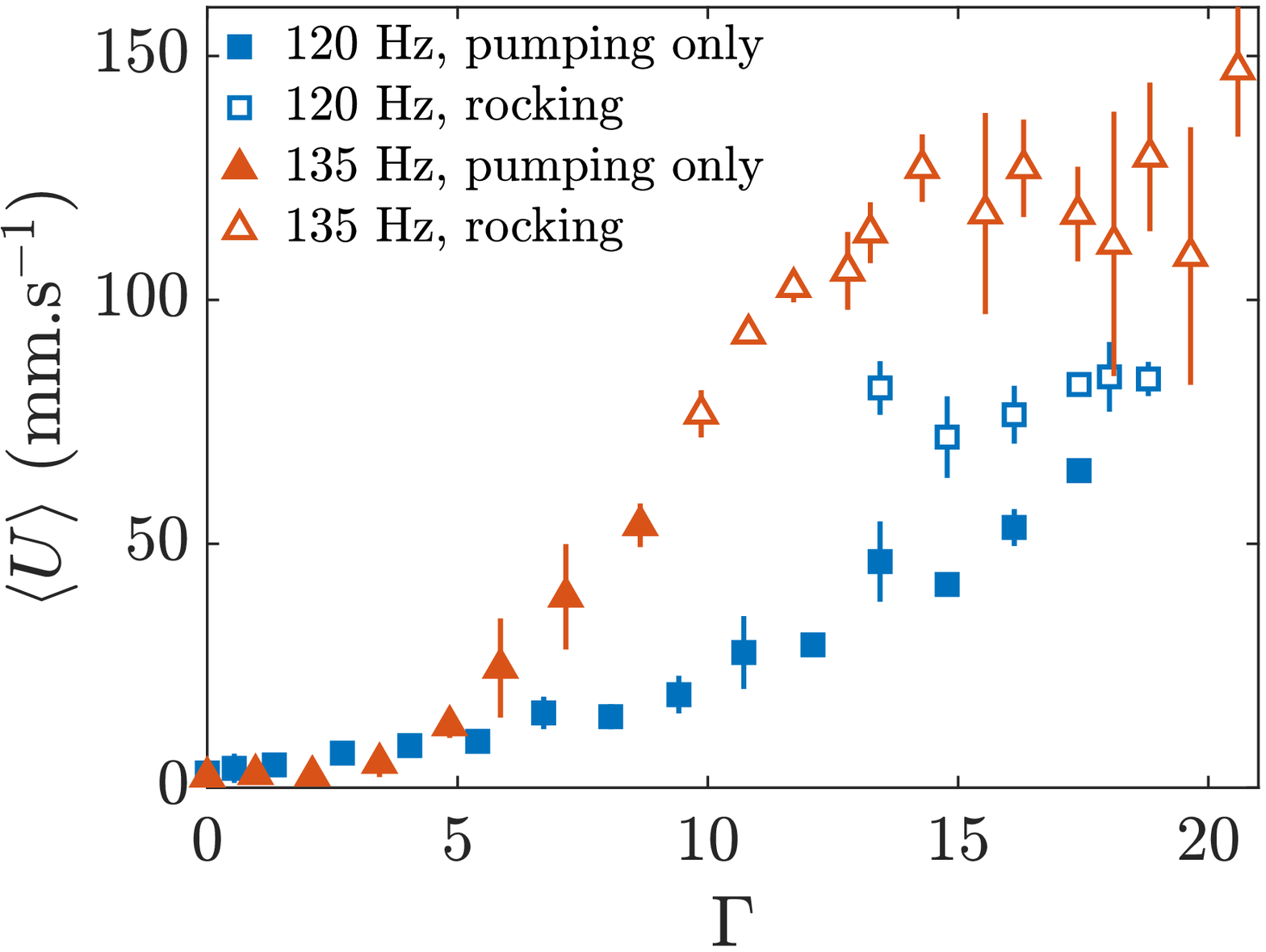}};
    	\node[] at (fig.north west){$(d)$};
  	\end{tikzpicture}

  	\begin{tikzpicture}
    	\draw (0, 0) node[inner sep=0] (fig) {\includegraphics[width=0.6\textwidth]{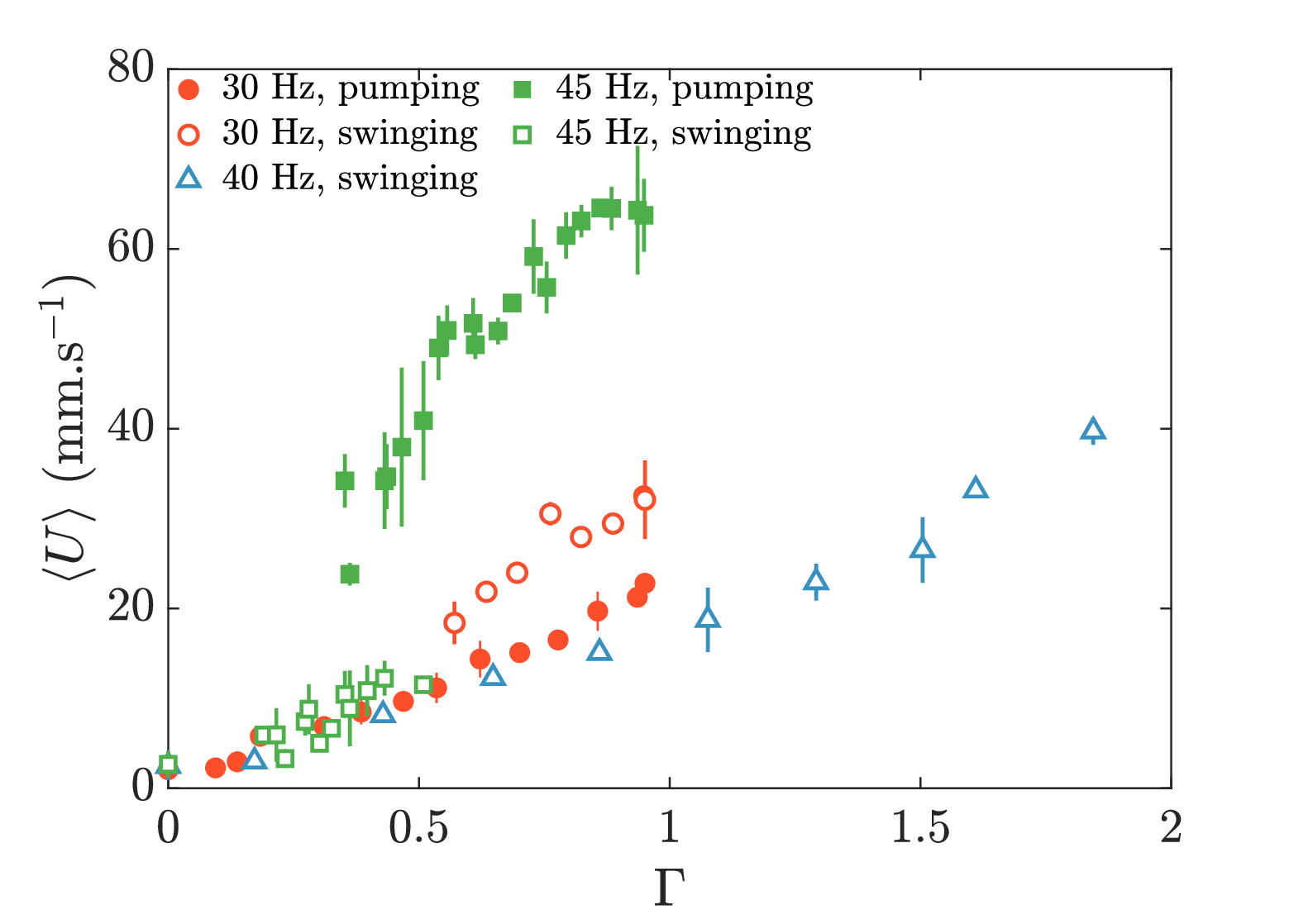}};
    	\node[] at (fig.north west){$(e)$};
  	\end{tikzpicture}
	\caption{Experimental results regarding the droplet's sliding speed $\langle U \rangle$ with  $V=4~\mu$L, $b=200~\mu$m, $\alpha=27.5^\circ$.
	$(a)$ Mobility parameter $s={\rm d}\langle U \rangle/{\rm d}A$ obtain from a linear fit  as a function of the forcing frequency $f$ considering only droplets responding with a harmonic pumping mode.
	$(b-e)$ $\langle U \rangle$ as a function of the normalized forcing acceleration $\Gamma$ for various frequencies $f$. Filled symbols represent experiments where the droplet exhibit harmonic pumping vibrations. Open symbols represent experiments where the droplet: $(c)$ exhibits subharmonic pumping vibrations at frequency $f/2$ ; $(d)$ exhibits both harmonic pumping and rocking modes ; $(e)$ swings subharmonically in a pendulum-like fashion at $f/2$.
	These various responses are illustrated in figures \ref{fig:subharmo}, \ref{fig:weird120} and \ref{fig:pendulum30}, respectively.
	We note that we have chosen to represent $\langle U \rangle$ as a function of the normalized acceleration $\Gamma=A\omega^2/g$ rather than as a function of the amplitude $A$: this  allows to compare more easily data with different frequencies.
The resonant response shown in $(a)$ is also evident when representing the averaged slope ${\rm d}\langle U \rangle/{\rm d}\gamma$ as a function of $f$ (not shown).
	\label{fig:speed_frequency}}
\end{figure}

Our main dataset focuses on a nylon fibre of diameter $b=200~\mu$m  making an angle $\alpha=27.5^\circ$ with the horizontal, on which a water droplet of volume  $V=4~\mu$L is deposited.
We systematically vary the frequency of oscillations from $f=15$ to 135 Hz, and the amplitude from zero and up to the detachment of the droplet from the fibre.
Reported values of the speed of the droplet $\langle U \rangle$ shown in figures \ref{fig:speed_frequency}, \ref{fig:fibre_diameter} and  \ref{fig:fibre_angle} represent the mean and standard deviation of typically 3 different experiments.
We note that with these parameters the droplet naturally slides down the fibre with a speed $U_0=2.5 \pm 1.6$ m.s$^{-1}$, the droplet's speed without vibrations ($A=0,~\Gamma=0$).
We will see next that there is no significant qualitative change in our observations when the droplet is pinned on the fibre for lower angles $\alpha$ or large diameters $b$ when the fibre does not oscillate ($U_0=0$).

As we discussed in the Introduction \S \ref{sec:Intro}, an empirical correlation between the forcing amplitude and the droplet's speed is given by \eqref{eq:dropmotion}, which is typically linear ($\chi=1$) in most experimental work using flat substrates.
We therefore expect a linear relationship between the droplet sliding speed and the amplitude of vibrations as $\langle U \rangle-U_0 = s(A-A_{\rm th})$ for $A>A_{\rm th}$, which allows us to define $s={\rm d}\langle U \rangle/{\rm d}A$.
In figure \ref{fig:speed_frequency}$(a)$ we show the evolution of $s$ as a function of the forcing frequency $f$: we observe a resonant behavior with a maximum for $f \approx 50$ Hz, which is near but slightly below the natural pumping frequency $f_{\rm pump}=57$ Hz discussed in \S\ref{subsec:oscillationfreq}.
We note that \citet{Costalonga2020}, using a different setup, found a maximum of mobility $s$ for a frequency close to, but in their case larger than, $f_{\rm pump}$.
Figure \ref{fig:speed_frequency}$(b)$ shows $\langle U \rangle$ as a function of $\Gamma$ for selected frequencies where the linear relation  \eqref{eq:dropmotion} with $\chi=1$ indeed seems to be satisfactory. 

In figure \ref{fig:speed_frequency}$(a)$ we extract $s$ only considering droplets that respond solely with harmonic pumping.
This mode of response is shown in figure \ref{fig:setup}$(b)$ and corresponds to droplets that periodically stretch and flatten with the same frequency as the forcing frequency $f$.
The associated data is represented with filled markers in figure \ref{fig:speed_frequency}$(b-e)$;
we will discuss in the next section the other regimes we have observed.
When the relationship between the droplet's speed and the forcing amplitude is nonlinear, we extracted $s$ from a linear fit but for small amplitudes only in order to compare with the other datasets.
Indeed while the exponent $\chi =1$ is reasonable for most frequencies, some of the data would be fitted more adequately with $\chi > 1$, e.g. $f=15$ Hz in figure \ref{fig:speed_frequency}$(b)$.
Such superlinear behaviour is common in numerical studies and has also been observed in some of the experiments of \citet{Costalonga2020}.
In Appendix $\ref{appendix:extra}$  we also show data for a larger  fibre diameter $b=400~\mu$m (figure \ref{fig:fibre_diameter}) and smaller tilt angle $\alpha=15$ and $7.5^\circ$ (figure \ref{fig:fibre_angle}): in these cases $U_0=0$, and we still observe a monotonic increase of the sliding speed as a function of the amplitude of vibrations for harmonically pumping droplets. Varying the tilt angle still yields $\chi \approx 1$, while increasing the fibre diameter gives more consistently a sublinear behaviour with $\chi<1$.

\begin{figure}
	 \centering
	\begin{tikzpicture}
    	\draw (0, 0) node[inner sep=0] (fig) {\includegraphics[width=0.7\textwidth]{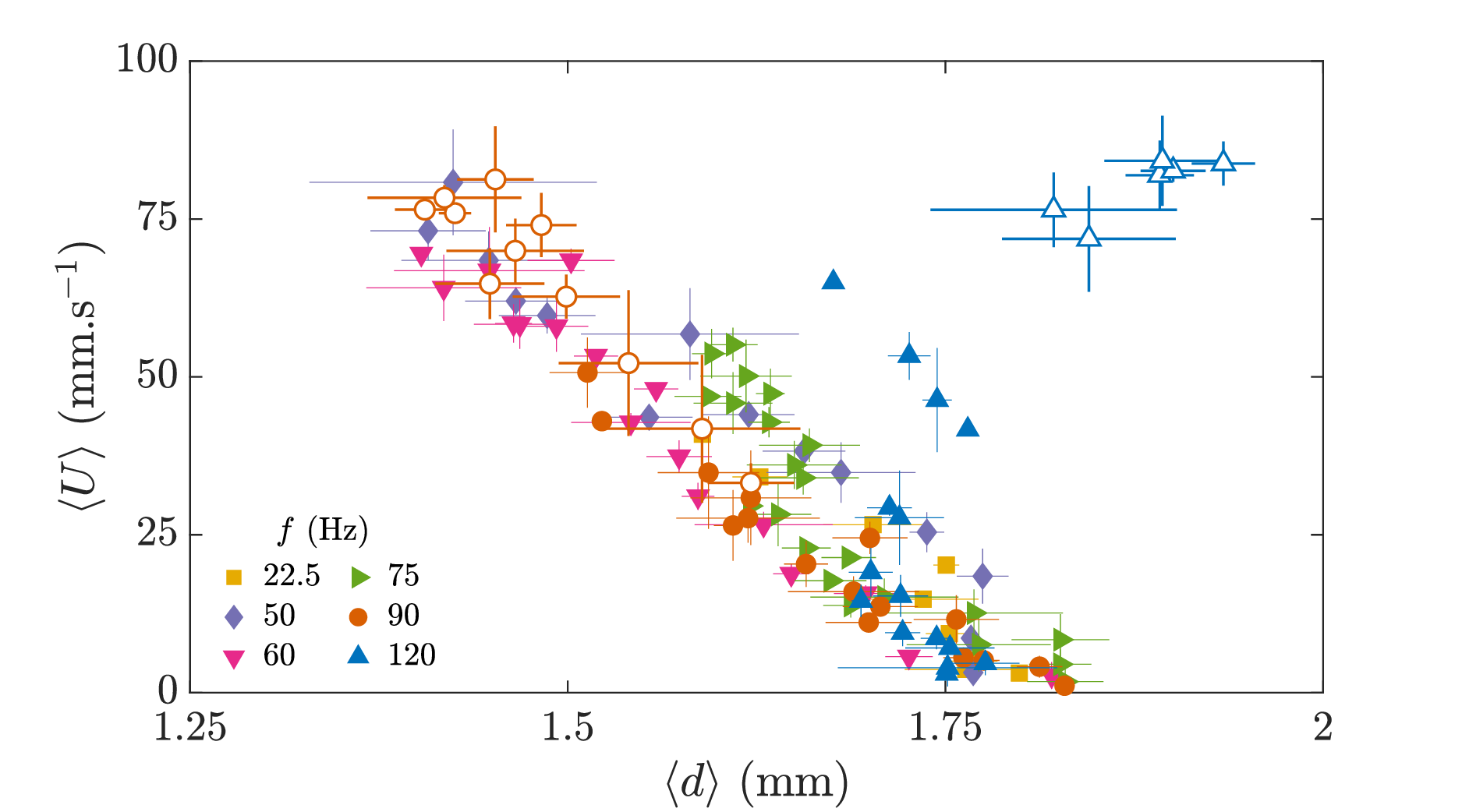}};
    	\node[] at (fig.north west){$(a)$};
  	\end{tikzpicture}

	\begin{tikzpicture}
    	\draw (0, 0) node[inner sep=0] (fig) {\includegraphics[width=0.47\textwidth]{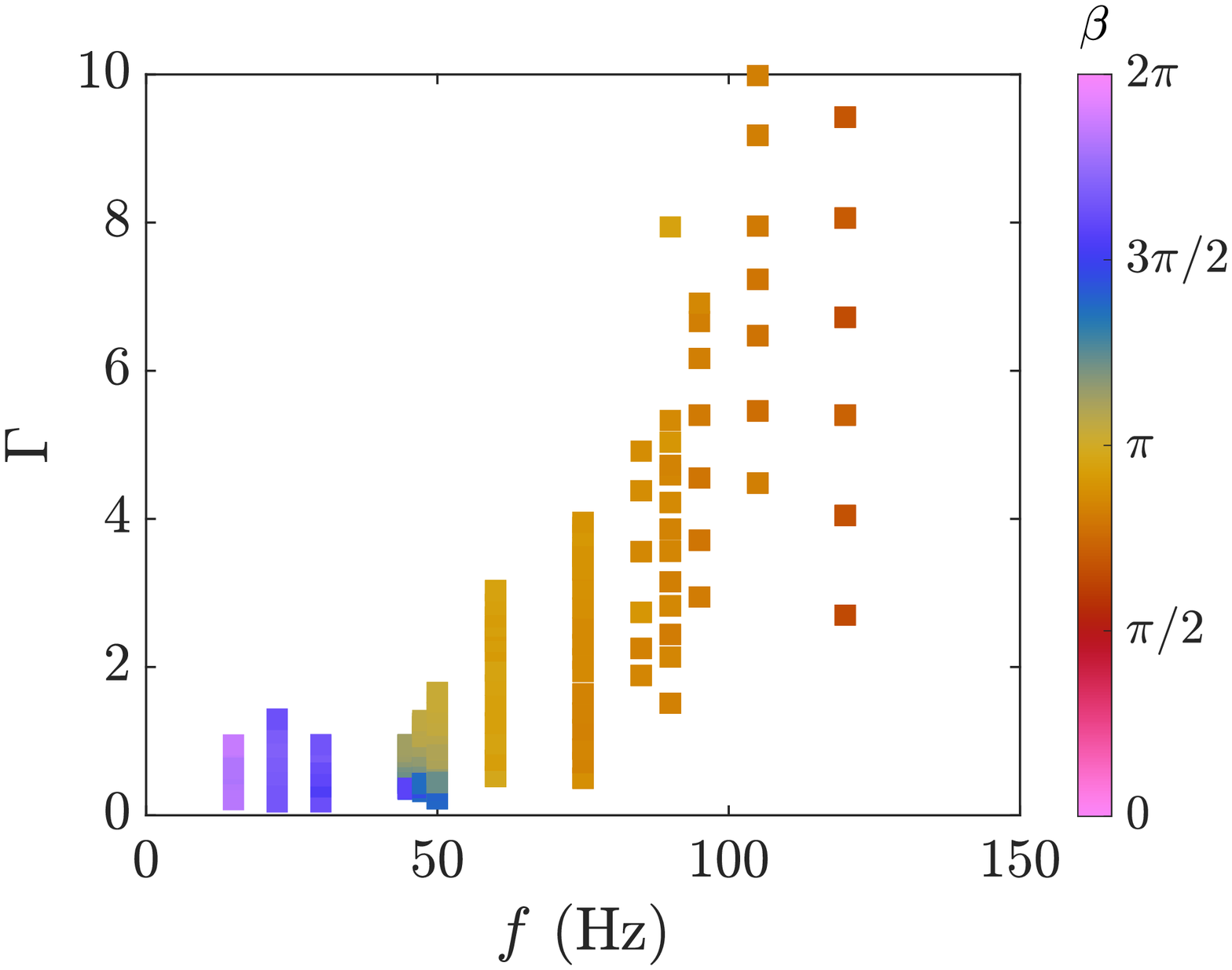}};
    	\node[] at (fig.north west){$(b)$};
	\end{tikzpicture}
	\begin{tikzpicture}
    	\draw (0, 0) node[inner sep=0] (fig) {\includegraphics[width=0.47\textwidth]{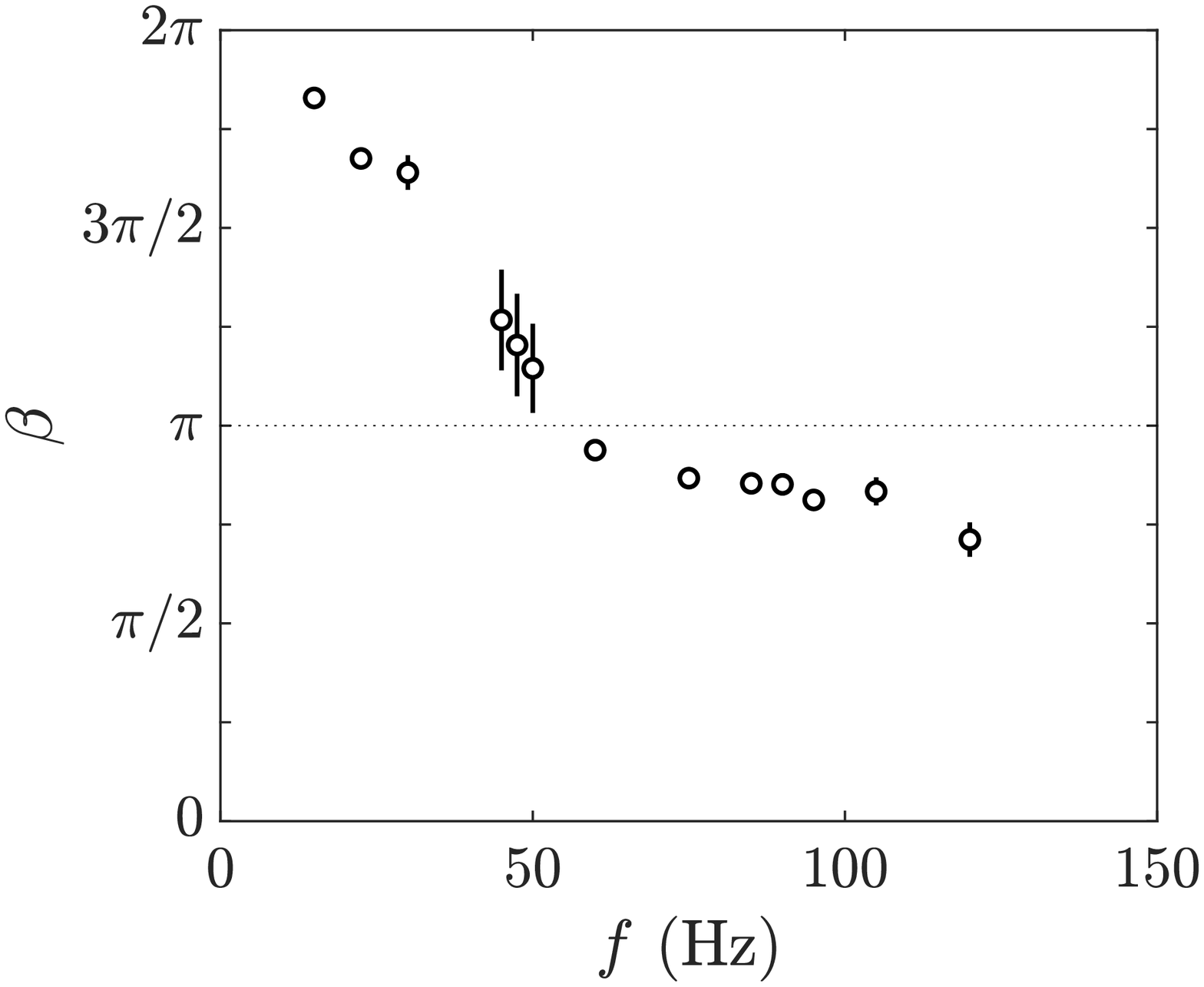}};
    	\node[] at (fig.north west){$(c)$};
	\end{tikzpicture}

	\caption{Data corresponding to $V=4~\mu$L, $b=200~\mu$m, $\alpha=27.5^\circ$.
	$(a)$ Correlation between the droplet's sliding speed $\langle U \rangle$ and the averaged basal diameter $\langle d \rangle=\int_0^T d(t)~{\rm d}t/T$ for various frequencies $f$  upon varying the amplitude of fibre's oscillations.
	Filled symbols represent data where the droplets exhibits a harmonic pumping mode.
	Open circles ($\circ$) for $f=90$ Hz correspond to subharmonic pumping while open upwards triangle ($\triangle$) for $f=120$ Hz correspond to a combination of pumping and rocking modes.
	$(b)$ Heatmap of the phase angle $\beta$ between the basal diameter $d$ and the fibre's position $y_{\rm fibre}$ for various frequencies and amplitudes of vibrations. This phase angle is computed using a Fourier analysis of the two signals $d(t)$ and $y_{\rm fibre}(t)$.
	Only droplets showing a harmonic pumping mode are considered.
	$(c)$ Mean value of $\beta$ for each frequency, averaged over all the amplitudes.
	In order to account for the fact that an angle is defined modulo $2\pi$, the mean value and standard deviation (used as errorbars) of $\beta$ are defined as the following directional moments: $\arg(m)$ and $(-2\ln(\lvert m\rvert))^{1/2}$, respectively, with $m=\sum_{j=1}^n \exp(i\beta_j)/n$.
	\label{fig:diameter}}
\end{figure}

Harmonic pumping vibrations of the droplet modulate its basal diameter $d$ at the forcing frequency $f$.
To investigate the trend between the droplet's speed and the amplitude of oscillations, we first consider the average diameter $\langle d \rangle=(1/T)\int_0^T d(t)~{\rm d}t$.
Figure \ref{fig:diameter}$(a)$ shows a correlation between $\langle d \rangle$ and the average droplet's sliding speed $\langle U \rangle$.
Despite some scatter in the data, a clear correlation emerges: for a given frequency, as the amplitude of fibre's oscillations increases, $\langle d \rangle$ decreases while  $\langle U \rangle$ increases. 
We note that given the shape of the droplet, $d$ is proportional to the wetted area of the droplet on the fibre $S\simeq \pi b d$.
It is also interesting to look at the phase angle $\beta$ between the basal diameter $d$ and the fibre position $y_{\rm fibre}$: $\beta=0$ or $2\pi$ corresponds to an evolution where $d$ is maximal at the crest of the fibre's oscillations, while $\beta=\pi$ corresponds to the opposite situation where $d$ is minimal at the crest. Figure \ref{fig:setup}$(b,c)$ shows an example where $\beta \approx \pi$.
Figure \ref{fig:diameter}$(b)$ shows $\beta$ for various frequencies and amplitudes of oscillations, where it appears that for a fixed frequency there is generally little change of $\beta$ upon varying the amplitude.
We show the evolution of $\beta$ averaged over all amplitudes as a function of the frequency $f$ in figure \ref{fig:diameter}$(c)$:
for $f\lesssim50$ Hz , $\beta \gtrsim 3\pi/2$, while it drops to  $\beta \approx \pi$  for $f\gtrsim 50$ Hz.
In fact this phase shift already occurs for 45 Hz $\leq f \leq$ 50 Hz upon increasing the amplitude of oscillations.
We note that $\beta$ has  been correlated with the speed and mobility of droplets in prior works on vibrating flat substrate.
\citet{Sartori2019} delimited regimes of descending and fast descending droplets, where $\beta \approx \pi$ in the descending regime and $\beta$ close to 0, or $2\pi$, in the fast descending regime.
Similarly \citet{Costalonga2020} found $\beta \approx 0$ near the maximum droplet mobility $s$, while $\beta \approx \pi$ corresponds to climbing drops.

The discussion above only focused on droplets responding to the fibre's oscillations with a harmonic pumping motion, but this is not the only interfacial motion that is produced.
While figures \ref{fig:setup}$(b,c)$ and \ref{fig:speed_frequency}$(a,b)$ summarize some of our observations on the droplet speed as a function of the amplitude, it also hides some complex interfacial flows: we have observed different regimes of droplet's response for $f\approx 90$ Hz, $f\geq 120$ Hz, and $30 \lesssim f\lesssim 45$ Hz.
We show in figure \ref{fig:speed_frequency}$(c-e)$ the effects that these different regimes have on the sliding speed and focus next on these.

\section{Regimes of droplet response}
\label{sec:regimes}
For some forcing frequencies the droplet can transition from one regime of vibrations to another upon increasing the forcing amplitude, with important effects on the sliding speed. We now discuss these different transitions in turn.

\subsection{Transition from harmonic to subharmonic pumping for $f \approx 90~{\rm Hz}$}
\label{subsec:subharmo}
\begin{figure}
	 \centering
	\includegraphics[width=\linewidth]{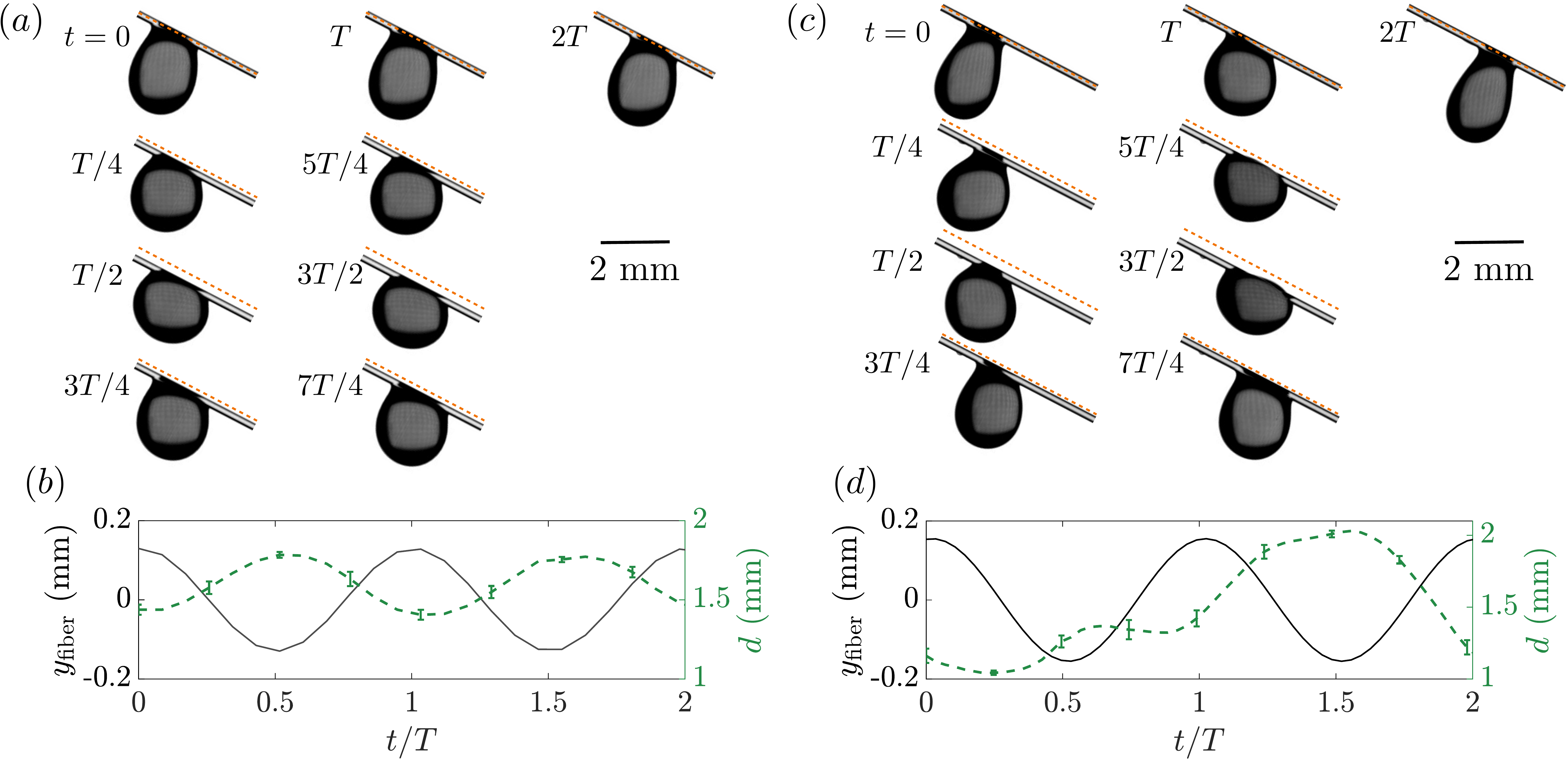}
	\caption{Illustration of the harmonic and subharmonic droplet behaviors observed at $f=90$ Hz with  $V=4~\mu$L, $b=200~\mu$m, $\alpha=27.5^\circ$ and $(a,b)$ $A=0.14$ mm, $\Gamma=4.7$ $(c,d)$ $A=0.17$ mm, $\Gamma=5.5$.
	$(a,c)$ Snapshots showing that the shape of the droplet is periodic with period $T$ and $2T$, respectively.
	The dotted line is fixed in the laboratory frame and represents the maximum position of the fibre, it highlights its oscillations.
	$(b,d)$ Corresponding evolution of the fibre's position (solid line) and basal diameter (dashed line).
	The mean value of the droplet's speed is $(a,b)$ $\langle U \rangle=31$ mm.s$^{-1}$ and $(c,d)$ $\langle U \rangle=58$ mm.s$^{-1}$. See movies 1 and 2 available in the supplementary movies.
	\label{fig:subharmo}}
\end{figure}

Figure \ref{fig:speed_frequency}$(c)$ shows $\langle U \rangle$ as a function of $\Gamma$ for $f=90$ and 95 Hz.
Above a threshold normalized acceleration $\Gamma_{\rm sub} \simeq 5$ and $6$, respectively, we observe that the droplet transitions from a regime of harmonic pumping to a regime of subharmonic pumping, where the droplet responses at half the forcing frequency $f$. We note that we also observed the same behavior for $f=85$ Hz, not shown for clarity since the corresponding data is very close to those with $f=90$ Hz.
This transition from harmonic to subharmonic response corresponds to a sharp increase of the sliding speed $\langle U \rangle$ for $f=85$ and 90 Hz. There is also an increase, albeit more moderate, at $f=95$ Hz.

Figure \ref{fig:subharmo} shows the difference in shapes and dynamics of two representative experiments performed at $f=90$ Hz and near $\Gamma_{\rm sub}$, with a harmonic response for $\Gamma=4.7<\Gamma_{\rm sub}$ and a subharmonic response for $\Gamma=5.5>\Gamma_{\rm sub}$, respectively.
Despite a relatively small change in the forcing amplitude, we observe a doubling of the speed of the droplet's center of mass speed $\langle U \rangle$.
Interestingly, one of the most obvious difference between the two droplets is regarding their basal diameter $d$. It shows little variation in the harmonic regime, evolving from 1.4 to 1.8 mm, compared to the subharmonic region, when it goes down to 1 and up to 2 mm. Its averaged value $\langle d \rangle$ is also smaller in the subharmonic case, and in fact figure \ref{fig:diameter}$(a)$ shows that the correlation between $\langle d \rangle$ and $\langle U \rangle$ previously discussed still holds.
In appendix \ref{appendix:extra} we show that this subharmonic response appears and also causes a jump in speed for thicker fibres or smaller tilt angles.

It is interesting to put these observations in perspective with prior work on droplets moving on a flat substrate.
First, the transition from a harmonic to a subharmonic behavior above a threshold forcing is reminiscent of a parametric instability. \citet{Costalonga2020} also observed experimentally the possibility of subharmonic response in their setup of droplets on a horizontal substrate submitted to slanted vibrations.
However, they report a transition from harmonic response for sliding droplets ($\langle U \rangle>0$) to subharmonic response for climbing droplets ($\langle U \rangle<0$), while we observe an acceleration of the descending speed in the subharmonic regime.
It is also interesting to mention that they obtained a subharmonic regime for $f\simeq 1.5f_{\rm pump}$; this 1.5 factor also match our experiments ($1.5 f_{\rm pump} \simeq 85$ Hz).
Second, using droplets on tilted liquid infused substrates submitted to vertical vibrations, \citet{Sartori2019} observed experimentally above a threshold acceleration a regime that they refer to as fast descending, where droplets slide much faster.
This regime associated with a basal diameter showing much more important variations than in the regular descending regime, similarly to our experiments.
However, \citet{Sartori2019} correlate the transition from  descending to fast descending to a switch of phase between $d$ and $y_{\rm fibre}$, but in both cases the droplet keeps a harmonic motion. 
Finally, through numerical simulations, \citet{Ding2018} reproduced the experiments of \citet{Brunet2007} of droplets on tilted substrates with vertical vibrations. 
Their results suggest the strong  importance of a non-sinusoidal evolution of the wetted area $S\approx\pi b d$.
We also see in figure \ref{fig:subharmo}$(b,d)$ that $d$ switches from near-sinusoidal in the harmonic case to completely non-sinusoidal when the response is subharmonic.

\subsection{Transition from pumping to rocking for $f \geq 120~{\rm Hz}$}
\label{subsec:rocking}

\begin{figure}
	 \centering
	\includegraphics[width=0.84\linewidth]{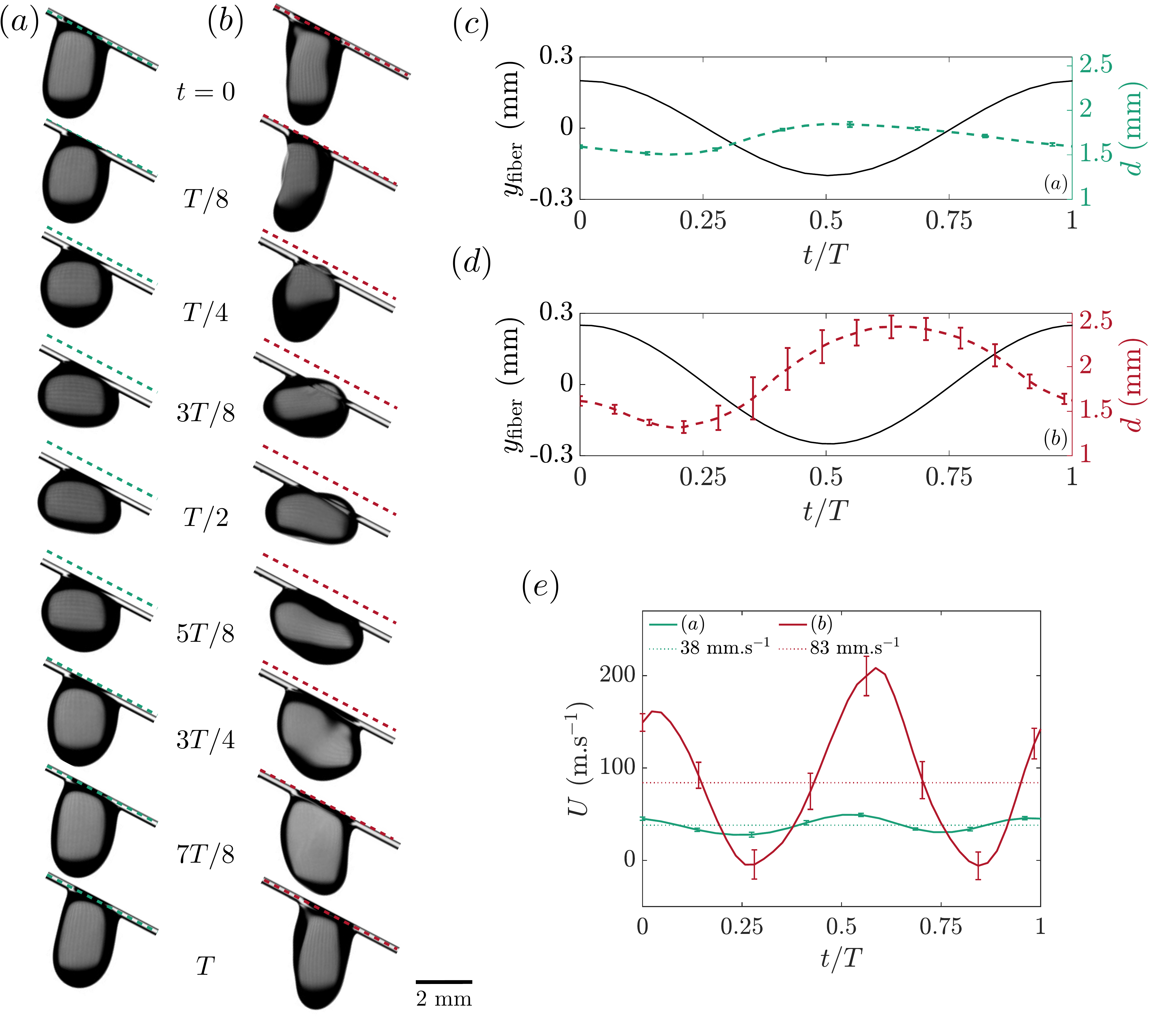}
	\caption{Droplets with $f=120$ Hz,  $V=4~\mu$L, $b=200~\mu m$, $\alpha=27.5^\circ$ and  $(a,b)$ $A=0.23$ mm, $\Gamma=13.4$ $(c,d)$ $A=0.31$ mm, $\Gamma=18.0$.
	The difference in the shape of the droplet over one period is illustrated in $(a,b)$.
	The dotted line represents the maximum value of $y_{\rm fibre}$ and highlights the fibre's vibrations.
	$(c)$ and $(d)$ show the time-evolution of the basal diameter for the droplets in $(a)$ and $(b)$, respectively.
	$(e)$ Instantaneous droplet speed. Averaged over one period, the mean value is $(a)$ $\langle U \rangle=38$ mm.s$^{-1}$ and $(b)$ $\langle U \rangle=83$ mm.s$^{-1}$. See movies 3 and 4 available in the supplementary movies.
	\label{fig:weird120}}
\end{figure}

Most droplets we have observed only exhibit a pumping mode, shown already in figures \ref{fig:setup}$(b,c)$ and \ref{fig:subharmo}.
For $f=120$ and 135 Hz and for high enough amplitude of vibrations, we observe a transition where the droplet can exhibit a combination of pumping and rocking modes. This is illustrated in figure \ref{fig:weird120}.
The existence of this rocking mode is particularly evident when considering the instantaneous speed ${\rm d}x_{\rm drop}/{\rm d}t$, figure \ref{fig:weird120}$(e)$.
In the pumping-only mode the droplet exhibits a near-constant velocity, showing variations of $\approx 20\%$ around the mean value $\langle U \rangle$: this is because pumping vibrations are mostly normal to the fibre. When the rocking mode appears, lateral vibrations become significant and the speed of the center of mass of the droplet shows significant variations around the mean. 

As shown in figure \ref{fig:speed_frequency}($d$), this rocking mode significantly increases the sliding speed $\langle U \rangle$.
This happens despite the fact that the averaged basal diameter $\langle d \rangle$ is on average larger in the presence of the rocking mode (figure \ref{fig:weird120}$c,d$), and the correlation of figure \ref{fig:diameter} does not hold anymore.
In this regime the combined effects rocking and pumping effects cannot be captured solely by the change in wetted area, similarly to what is observed on droplets on horizontal flat surfaces submitted to slanted vibrations (see \cite{Costalonga2020} and the discussion in \S\ref{sec:Intro}).

\subsection{Transition between pumping and swinging for $30 {~\rm Hz} \lesssim f \lesssim 45 {~\rm Hz}$}
\label{subsec:swinging}

\begin{figure}
	 \centering
	\includegraphics[width=0.9\linewidth]{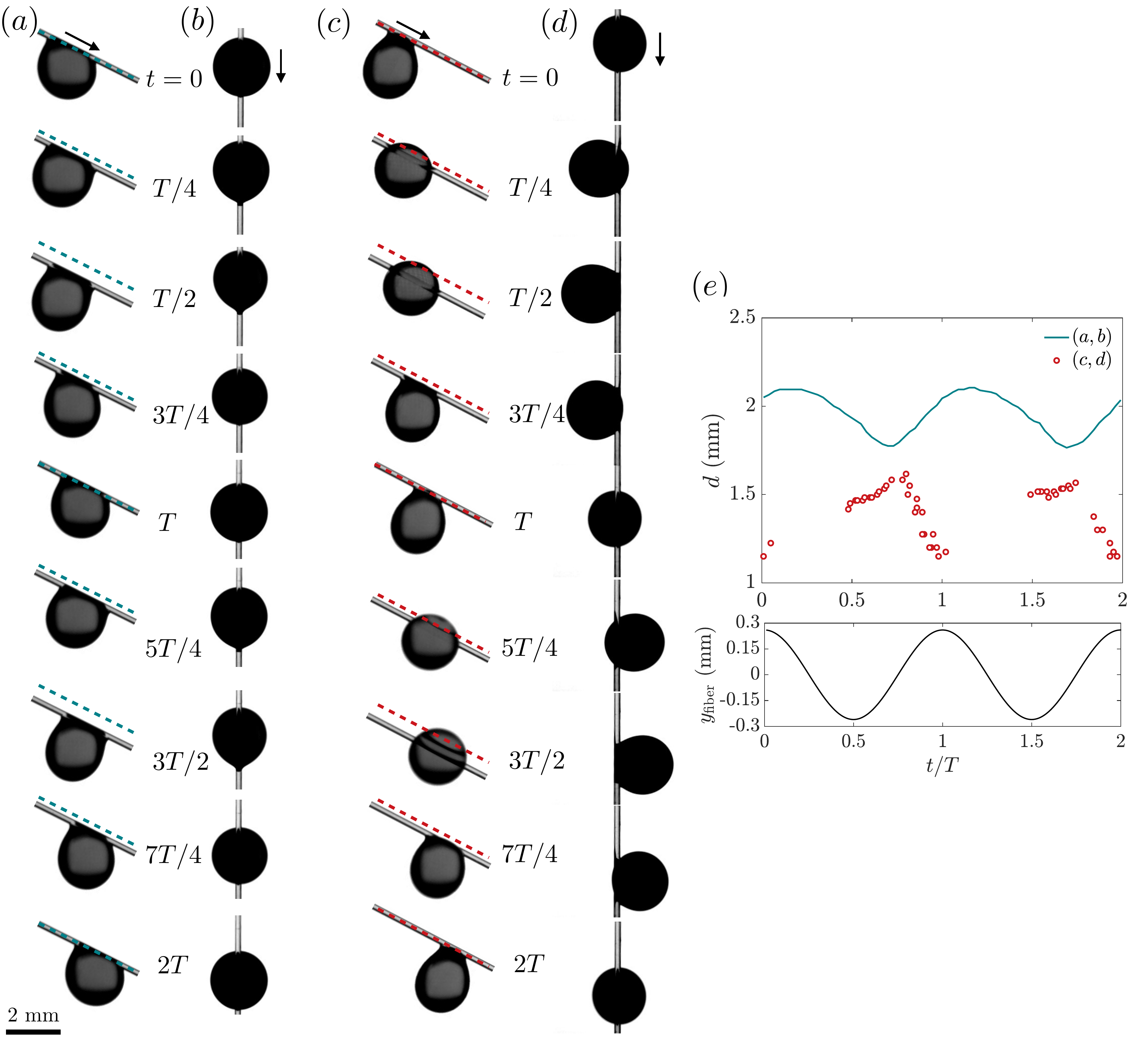}
	\caption{Droplets at $f=30$ Hz  $V=4~\mu$L, $b=200~\mu$m, $\alpha=27.5^\circ$ and $A=0.26$ mm, $\Gamma=0.96$.
	 $(a,b)$ shows a vibrating-only regime where the droplet slides at $\langle U \rangle=18$ mm.s$^{-1}$,
	 $(c,d)$ shows a subharmonic, swinging regime where the droplet slides at $\langle U \rangle=31$ mm.s$^{-1}$.
	 $(a,c)$ are viewed from the side and $(b,d)$  from the top, looking down at the droplet.
	 $(e)$  Time evolution of the basal diameter $d$ extracted from the videos corresponding to $(a,b)$ and $(c,d)$. In the latter case, when the droplet swings, we  did not extract $d$ with automatic image processing but did manual measurements combining both the side and top views as needed. We could not extract accurate data for $0<t<0.5T$ and $T<t<1.5T$.
	 See movies 5 to 8 available in the supplementary movies.
	\label{fig:pendulum30}}
\end{figure}

Figure \ref{fig:speed_frequency}$(e)$ shows the speed of droplets as a function of the amplitude of oscillations for $f=30,~40$ and 45 Hz.
For these three frequencies the droplet can respond by swinging across the fibre similarly to a pendulum as illustrated in figure \ref{fig:pendulum30}.
This swinging motion is subharmonic at half the forcing frequency.

At $f=30$ Hz, the swinging motion is only observed for high enough amplitude of oscillations and when droplets are significantly perturbed when, or after, being deposited on the fibre (e.g. due to the detachment from the micropipette, or by flicking the oscillating structure). When they are gently deposited on a still fibre with a slowly increasing amplitude of oscillations, only the harmonic pumping response is observed.
However once they enter the swinging mode they do not return to harmonic pumping. Henceforth for $f=30~{\rm Hz}$ the harmonic pumping response is unstable to finite perturbations.
Figure \ref{fig:speed_frequency}$(e)$ shows that the transition to the swinging mode significantly increases the sliding speed.

Figure \ref{fig:pendulum30}$(e)$ shows that in the swinging mode, the basal diameter $d$ and hence the wetting area $S$ are significantly smaller than in the pumping case. This can be understood at least partly by considering the centrifugal acceleration induced by the swinging motion.
The droplet has an equivalent spherical radius $r=(3V/4\pi)^{1/3} \simeq 1$ mm and swings at $f/2$ so that its angular velocity can be approximated, on average, as $\omega/2$.
The resulting centrifugal acceleration is $r\omega/2 \approx 9$ m.s$^{-2}$, which is comparable to both the gravitational acceleration and to the acceleration induced by the fibre motion (here $\Gamma \simeq 1$ and hence $A\omega^2 \simeq g=9.8$ m.s$^{-2}$). 
This supports the idea that the centrifugal acceleration due to swinging pushes the droplet away from the fibre, diminishing its wetting area.
However the data would not collapse on figure \ref{fig:diameter}$(a)$ showing the correlation between $\langle d \rangle$ and $\langle U \rangle$; this is not surprising and shows that energy is dissipated differently 
as the flow inside the droplet is very different when swinging as compared to when it is pumping.

When increasing the frequency to $f=40$ Hz, harmonic pumping becomes more and more unstable and eventually cannot be reached anymore.
Here we only observe subharmonic swinging even at low amplitude of oscillations.
Increasing again the frequency to $f=45$ Hz, we also only observed subharmonic swinging at low amplitude. 
However, upon increasing the amplitude the droplet switches to a subharmonic pumping mode.
At this frequency ($f=45$ Hz) it is now the swinging mode that becomes unstable at high amplitude; this transition also increases drastically the sliding speed (figure \ref{fig:speed_frequency}$e$).
Data with a different fibre diameter and different tilt angles shown in Appendix \ref{appendix:extra} also exhibit this behaviour.

While the two previously discussed regimes of rocking and subharmonic pumping are also observed on flat vibrating substrates, swinging droplets can only occur on fibres.
Next we aim to rationalize the existence of such swinging drops.

\section{Droplet swinging off a horizontal fibre{}}
\label{sec:swinging}
\subsection{Experiments}

\begin{figure}
	\centering
	\includegraphics[width=0.8\textwidth]{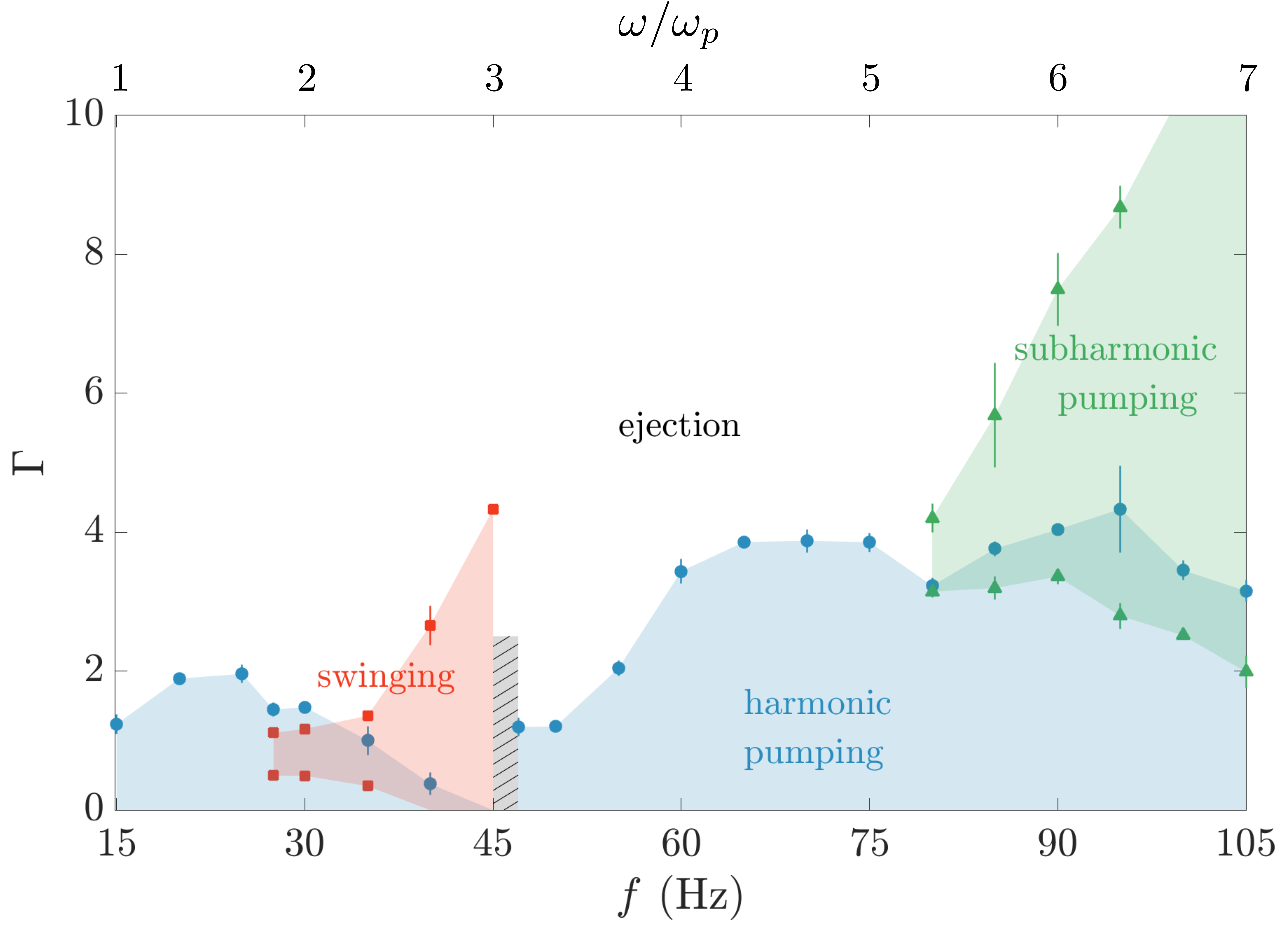}
	\caption{Regime map of the three different droplet behaviours ($V=4~\mu$L) on a horizontal fibre ($\alpha=0^\circ$) of thickness $b=200~\mu$m. The region between the horizontal line $\Gamma=0$ and the circles, in blue, represents the regime of harmonic vibrations.
	The region between triangles, in green, represents the regime of subharmonic vibrations at $f/2$.
	The region between squares, in red, represents the regime of subharmonic swinging at $f/2$.
	The grayed hatched region between 45 and 47 Hz corresponds to a narrow range of frequencies where it is challenging to obtain reproducible results.
	The top axis represent the dimensionless frequency $\omega/\omega_p$, with $f_p=\omega_p/2\pi$ the pendulum frequency discussed in the text. We have used $f_p=15$ Hz.
	\label{fig:regime_map}
	}
\end{figure}

In order to obtain a clearer picture of the pendulum-like swinging droplet motion illustrated in figure \ref{fig:pendulum30}, we focus on droplets on horizontal fibres ($\alpha=0^\circ$) to decouple the droplet's sliding motion to its response to oscillations.
We construct a flow regime map from experiments, %with a deionized water droplet of volume $V=4\mu$L on a $b=200~\mu$m nylon fibre,
where we observe three different behaviours: harmonic vibrations, subharmonic vibrations, and swinging.
It is shown in figure \ref{fig:regime_map}, confirming the observations in \S \ref{sec:slidingdrops} on tilted fibres, namely the existence of a subharmonic swinging regime for $f$ ranging from approximately 30 Hz and up to 45 Hz, and a subharmonic vibration starting near $80$ Hz.

We note that figure \ref{fig:regime_map} shows a region where we only observe swinging droplets, but also a region where both a harmonic vibrating droplets and  subharmonic swinging droplets can coexist.
In the latter regime, subharmonic swinging only takes place if a finite perturbation is introduced in the system (e.g., blowing gently on the droplet, depositing the droplet on an already vibrating fibre or flicking the vibrating structure), or when decreasing the amplitude of vibrations from the region where only swinging motion occurs.
Once a droplet is in the subharmonic swinging mode it enters a stable state, and we never observed a droplet switching from a swinging to a non-swinging motion.
We also note that this region of nonlinear instability is bounded from above at $f \approx 30$ Hz: this is not because the droplet switches to harmonic vibrations upon increasing the amplitude, but because it detaches from the fibre when the swinging mode is excited at higher amplitudes.

\begin{figure}
	 \centering
	\includegraphics[width=\linewidth]{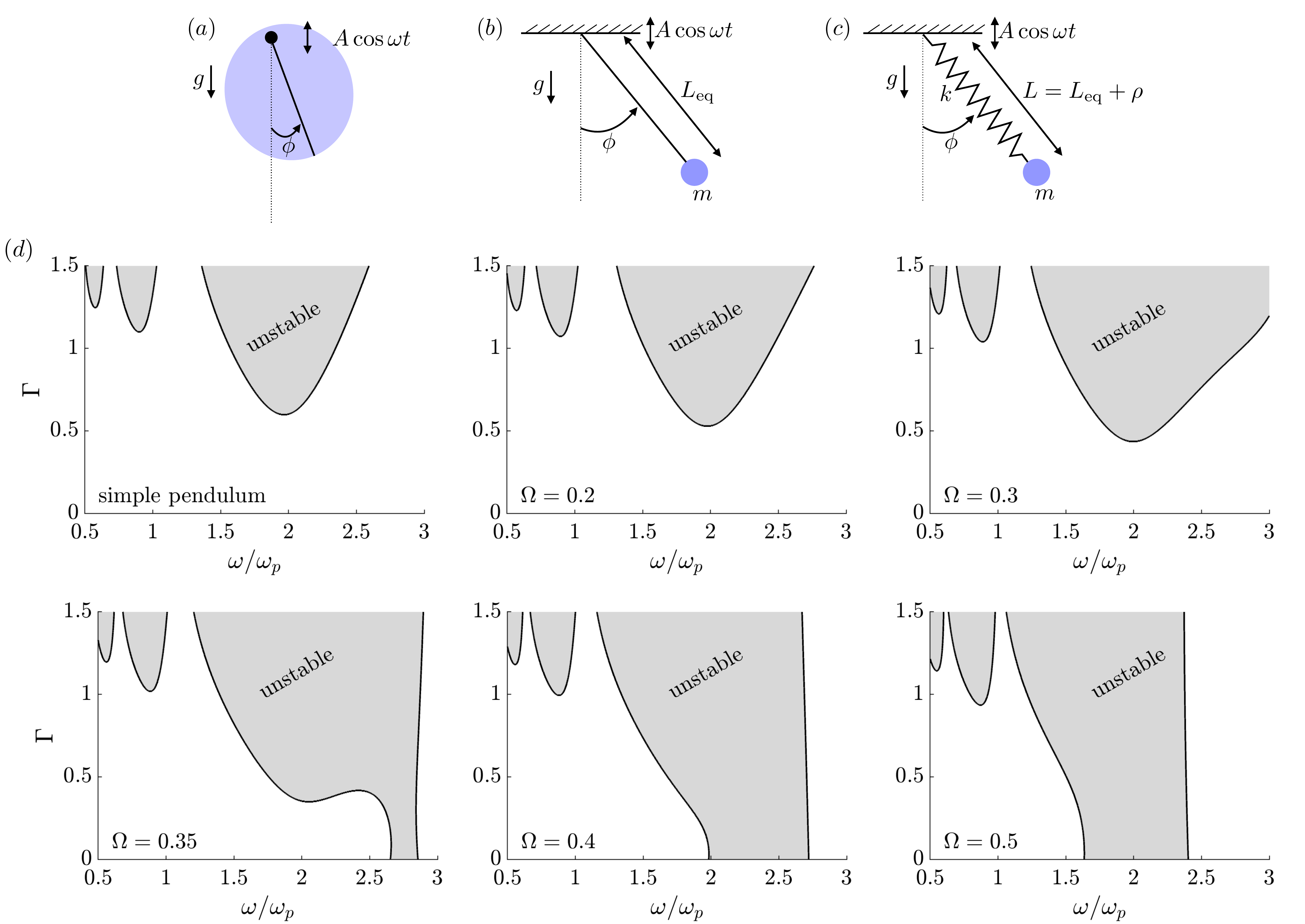}
	\caption{
	Analogy between $(a)$ a droplet swinging of a fibre -- visualized in a plane perpendicular to the cross section of the fibre -- and $(b)$ a simple pendulum or $(c)$ an elastic pendulum with spring constant $k$, with a point mass $m$ and an equilibrium length $L_{\rm eq}$. The system is placed in the gravitational field $g$, the support oscillates as $A\cos\omega t$, and $\phi$ denotes the angle with respect to the vertical.
	$(d)$ Linear stability diagrams of \eqref{eq:forcedMathieuModified} for a simple pendulum and an elastic pendulum for different values of $\Omega=\omega_p/\omega_s$, the ratio of the natural frequency of the pendulum motion $\omega_p=(g/L_{\rm eq})^{1/2}$ to that of the spring motion $\omega_s=(k/m)^{1/2}$. The damping coefficients are $c_s=c_p=0.15$, and $\Gamma=A\omega^2/g$ is the dimensionless forcing acceleration.
	\label{fig:pendulum}}
\end{figure}

\subsection{Analogy with a forced pendulum}
\label{subsec:forcedpendulum}
The swinging motion of the droplet is reminiscent of a pendulum.
We illustrate this analogy in figure \ref{fig:pendulum}$(a,b)$ between a droplet hanging below an oscillating fibre and a pendulum with constant length $L_{\rm eq}$, mass $m$, in the gravitational field $g$ and submitted to vertical oscillations of its support as $A\cos(\omega t)$.
By making this analogy we assume the droplet to be a solid with a motion described by a single degree of freedom, the angle $\phi$ with respect to the vertical direction, and ignore any interface deformations.
Such a forced pendulum obeys the following damped Mathieu equation \citep{Kovacic2018}, which is the classical pendulum equation written in the frame of the oscillating support:
\begin{align}
	\phi'' + 2{c_p}{\omega_p}\phi'+\omega_p^2\left(1-\Gamma\cos\left(\omega t\right)\right)\phi=0,\qquad \omega_p=(g/L_{\rm eq})^{1/2},
	\label{eq:Mathieu}
\end{align}
where  $(.)'={\rm d}(.)/{\rm d}t$ denotes the derivative with respect to time, and $\omega_p$ is the natural frequency of pendulum oscillations. 
We have assumed small angles with $\vert \phi \vert \ll 1$, and to account for dissipation we have included a linear damping term characterized by the dimensionless friction coefficient $c_p$.
Equation \eqref{eq:Mathieu} is an archetype for parametric instabilities with connections to a wide range of physical phenomena and has been studied extensively \citep{Kovacic2018}. 
The equilibrium position $\phi=0$ is unstable for large enough values of the forcing $\Gamma$ at driving frequencies $\omega\approx 2\omega_p/n$, with $n=1,2,3,\dots$ a positive integer. 
The most unstable mode corresponds to $n=1$, $\omega\simeq 2\omega_0$, where the pendulum then swings subharmonically at $\omega/2$.
A stability diagram of \eqref{eq:Mathieu} is shown in figure \ref{fig:pendulum}$(d)$ where the greyed regions represent the so-called Arnold tongues of instability, where the position $\phi=0$ is linearly unstable and the oscillations of the substrate make the pendulum swings.
The Arnold tongues would go down to $\Gamma=0$ in the undamped case ($c_p=0$), the main effect of damping is to shift the tongue of instabilities to higher values of $\Gamma$.% \citep{Turyn1993}.

We assume that the droplet hanging down a fibre can be mimicked as a simple pendulum and can then be modelled by \eqref{eq:Mathieu}, with $L_{\rm eq}=I/a$, $\omega_p=\left(mga/I\right)^{1/2}$ for a physical pendulum with a homogeneous distribution of mass.
 Here $a$ is the distance from the axis of rotation to the center of mass of the droplet and $I$ the moment of inertia of the droplet about the axis of rotation.
We estimate $a$ and $I$ from experimental images of a droplet on a static fibre ($\Gamma=0$) and find $f_p = \omega_p/2\pi \approx 15$ Hz for a \mbox{4 $\mu$L} water droplet on a $200~\mu$m horizontal nylon fibre.
From the discussion above we would then expect to observe a swinging motion centered around $f\approx2f_p=30~$Hz.
Experiments show that the range of unstable frequencies is \mbox{30 Hz $\lesssim f\lesssim$ 45 Hz}.

One explanation to this discrepancy could be an inaccurate estimation of the natural frequency $f_p$.
Another experimental procedure to measure it, distinct from the geometrical method described above, is to induce the swinging motion of the droplet on a still fibre and directly observe the period of oscillations.
 When perturbing the droplet using a mechanical impulse or blowing on it on a still fibre, we however did not manage to induce significant oscillations without detaching the droplet.
Instead we oscillated the fibre for the droplet to naturally swing, and suddenly stopped the oscillations.
The measured oscillation frequency is in this case $21$~Hz, which is different than the previous estimate $f_p=15$ Hz. 
This would predict a swinging motion with a tongue of instability centered around $21$ Hz, which might be considered in better agreement with our experimental results but still does not capture some of the experimental features, and in particular the fact that swinging cannot be observed above 45 Hz.
 With this method, however, the droplet is both swinging and vibrating in a pumping mode, and we cannot exclude an effect of the vibrations on the swinging itself. 
This is in fact always true and hints at the fact that the droplet pumping mode of response should be considered to explain the swinging region in the phase diagram shown in figure \ref{fig:regime_map}.

\subsection{Analogy with a forced elastic pendulum}
\label{subsec:forcedelasticpendulum}

To account for the vibrations of the droplet itself, in addition to the possibility of swinging, we add in the analogy discussed above a linear spring (figure \ref{fig:pendulum}$c$).
This spring allows for a modulation of the pendulum length, analogous to the pumping mode of vibration of the droplet that stretches and flattens it, and its spring constant $k$ plays the role of the surface tension coefficient.
This is a first-order model for the effect of surface tension and cannot capture all the features of droplet vibrations, but we use this minimal toy model to gain insights into the coupling between the swinging and pumping modes of vibrations.
In Appendix \ref{appendix:elasticpendulum} we show that for small angles $\phi$ the dynamics of an elastic pendulum is governed by the following
\begin{subequations}
\begin{align}
	&\rho(t)=\frac{-L_{\rm eq} \Omega^2}{\left(1-(\omega/\omega_s)^2\right)^2 + 4c_s^2(\omega/\omega_s)^2}\left[\Gamma\left(1-(\omega/\omega_s)^2\right)\cos(\omega t) \right.% \\
    \left. + 2c_s (\omega/\omega_s)\sin(\omega t) \right], \\
	&{\phi}'' + 2\left(c_p\omega_p +\frac{\rho'}{L_{\rm eq}+\rho}\right) {\phi}' + \omega_p^2 \frac{1-\Gamma\cos(\omega t)}{1+\rho/L_{\rm eq}}\phi = 0,
	\end{align}
		\label{eq:forcedMathieuModified}
\end{subequations}
with a one-way coupling between the vibrations characterized by the length $\rho$ and the swinging characterized by the angle $\phi$ (see figure \ref{fig:pendulum}$c$).
Here $\omega_s=(k/m)^{1/2}$ is the natural frequency of the spring, $c_s$ is a damping coefficient introduced to model viscous dissipation.
We define $\Omega=\omega_p/\omega_s$, the ratio between the natural pendulum frequency of the droplet to its natural frequency of vibrations.
For a droplet, we expect $\omega_s \sim (\sigma/\rho r^3)^{1/2}$ and $\omega_p\sim(g/r)^{1/2}$ with  $r\sim V^{1/3}$ a characteristic size of the droplet, so that $\Omega \sim {\rm Bo}^{1/2}$ with ${\rm Bo}=\rho g r^2/\sigma$ the Bond number that compares gravitational and capillary effects.

When $\rho/L_{\rm eq}$ remains small, \eqref{eq:forcedMathieuModified} simplifies to the damped Mathieu equation \eqref{eq:Mathieu} and the analysis discussed previously in \S \ref{subsec:forcedpendulum} applies. 
However when $\rho/L_{\rm eq}$ can vary significantly, the coefficients in \eqref{eq:forcedMathieuModified} remain periodic but depend on the evolution of $\rho(t)$.
This regime of significant variation of the pendulum length is expected to be  relevant close to the resonance frequency of the spring. 
We can then anticipate a possible effect on the instability diagram when the main unstable angular frequency of the pendulum $2\omega_p$ is close to the resonance frequency of the spring $\omega_s$, i.e., for $\Omega\approx 1/2$.

We have discussed experimental measurements of $\omega_p$ in \S \ref{subsec:forcedpendulum}, yielding $\omega_p \approx 15 - 22$ Hz.
In \S\ref{subsec:oscillationfreq} we discussed the natural frequency of the first pumping mode of the droplet: $f_{\rm pump} \simeq 57$ Hz. Taking $\omega_s \approx \omega_{\rm pump}$ this gives a ratio  $\Omega \approx 0.26 - 0.37$.
We fix the values of $c_p=c_s=0.15$ arbitrarily and compute numerically the linear stability diagrams of the equilibrium position $\phi=0$ in the $(\omega/\omega_p-\Gamma)$ plane for chosen values of $\Omega$.
We explain in Appendix \ref{appendix:elasticpendulum} how we have obtained these stability diagrams, and the results are shown in figure \ref{fig:pendulum} for $\Omega$ ranging from 0.2 to 0.5.
Additional data for different value of $c_p$ and $c_s$ are presented in figure \ref{fig:elasticpend_c}.
As expected, for $\Omega=0.2$ there is little influence of the spring on the stability of the swinging motion.
As $\Omega$ gets closer to 0.5 however, the modulation of the pendulum length due to the spring makes the position $\phi=0$ more easily unstable and lowers the main tongue of instability down to $\Gamma=0$, despite the presence of damping.
Interestingly, the high-frequency limit of the instability band near $\omega=2\omega_p$ becomes very steep: this feature is also observed in our experimental phase diagram shown in figure \ref{fig:regime_map} where the swinging motion is observed for all amplitudes at $f=45$ Hz, but is never reached at $47$ Hz and above.
This band is also not necessarily centered near $\omega = 2\omega_p$ but is shifted to higher values of $\omega$ for $\Omega<0.5$, this is consistent with the apparent underestimation of $\omega_p$ from geometric considerations.

\section{Discussion}

In this article we have described the behaviour of water droplets deposited on small titled fibres that undergo sinusoidal oscillations. 
For a given frequency of oscillations, the higher the amplitude is, the faster the droplet slides down the fibre (figure \ref{fig:speed_frequency}).
The relation between the amplitude of oscillations and the droplet's sliding speed is typically linear, i.e. $\chi = 1$ in \eqref{eq:dropmotion}, even though both sublinear and superlinear behaviours can also be observed.
Most of our observations are droplets which present a harmonic pumping motion, which periodically flattens and stretches the droplet, modulating its wetted area (figure \ref{fig:diameter}$a$). 
However, it is not the viscous stresses distributed over the wetted area that control the droplet's motion, but rather the viscous dissipation located near the contact line. 
Prior work on vibrating flat substrates by \citet{Brunet2007,Costalonga2020} successfully linked the droplet's speed to the distribution of contact angle along the contact line using the unbalanced Young's law. It is challenging to do see here given the fibre geometry that makes difficult to measure accurately contact angles.

Besides harmonic pumping, we observed three other possible responses of the droplet. At a fixed frequency this is associated transitions between the different regimes upon increasing the amplitude of oscillations:
i/from harmonic pumping to subharmonic pumping for $f\approx 90$ Hz (figure \ref{fig:subharmo});
ii/from harmonic pumping to a combination of harmonic pumping and rocking for $f\geq 120$ Hz (figure \ref{fig:weird120});
iii/from harmonic pumping to subharmonic swinging for $f=30$ Hz (figure \ref{fig:pendulum30});
and iv/from subharmonic swinging to harmonic pumping for $f=45$ Hz.
All of these transitions greatly increase the droplet's sliding speed (figure \ref{fig:speed_frequency}).

The first two regimes are not unexpected.
The combination between rocking and pumping modes is ubiquitous for droplets on flat substrates and is one of the main observed driver of directional motion.
Subharmonic responses are also ubiquitous in free surface flows and have been observed as well for drops on vibrating planes. 
The other two transitions associated with a swinging motion are due to the specific geometry of a droplet hanging down a fibre geometry.
We rationalized this swinging motion by developing a simple model with only two degrees of freedom: a mass attached by a linear spring to an oscillating base.
In this elastic pendulum analogy, the spring is a simple model for the modulation of the droplet's height that is resisted by surface tension.
This yields a modified Mathieu equation able to explain why a droplet pumping vertically below the fibre is unstable and starts to swing.
Despite being a minimal model, this explains why the swinging motion appears, why there is a narrow range of frequency where the swinging motion is excited even at very low amplitude of oscillations, and why there is a sharp cut-off frequency after which swinging is never observed even at high amplitudes.
However it is limited to droplets on horizontal fibres, which do not show exactly the same features as droplets on tilted fibres that in addition slide.
In particular we do not explain the fourth transition discussed above, where the swinging motion is suppressed upon increasing the amplitude of oscillations at 45 Hz.
%\todo{2D despite minimal model it can capture many features: strength}
More generally, and similarly to studies on flat substrates, it remains challenging to link the droplet's mode of response to its sliding speed.
Detailed characterization of the response of drops on horizontal oscillating fibres, similar to the extensive body of literature on drops on flat horizontal substrates (e.g. \citet{Bostwick2015,Chang2015}), would be very interesting to understand fully the changes induced by the fibre geometry alone.

Vibrations of drops on fibres have been reported both in the context of aerosol filtration \citep{Dawar2006,Dawar2008} and fog harvesting \cite{Zhang2018}, where it was observed to trigger or enhance motion.
Wind is in fact present in many situations involving drops and fibres and can induce structural oscillations; we demonstrated here the impact this can have on the transport of droplets on fibres.
In applications such as digital microfluidics, oscillations can be controlled to induce motion and precisely vary the droplet speed.
In addition to the shape and surface properties of the fibre, structural oscillations therefore have the potential to be a design parameter to control the behavior of drops.

\section*{Declaration of Interests}
The authors report no conflict of interest.

\section*{Acknowledgments}
We thank Olav Gundersen for help with the experimental setup, Dr. Annette Cazaubiel for useful discussions, Dr. Vanessa Kern for useful discussions and for performing the measurements of static contact angle, and Dr. L. Mahadevan for stimulating discussions about the Mathieu equation. 

\section*{Funding}
This research has been founded by the Research Council of Norway through the program NANO2021, project number 301138.

\appendix

\section{Additional data}
\label{appendix:extra}
\begin{figure}
	\centering
	\begin{tikzpicture}
    	\draw (0, 0) node[inner sep=0] (fig) {\includegraphics[width=0.47\textwidth]{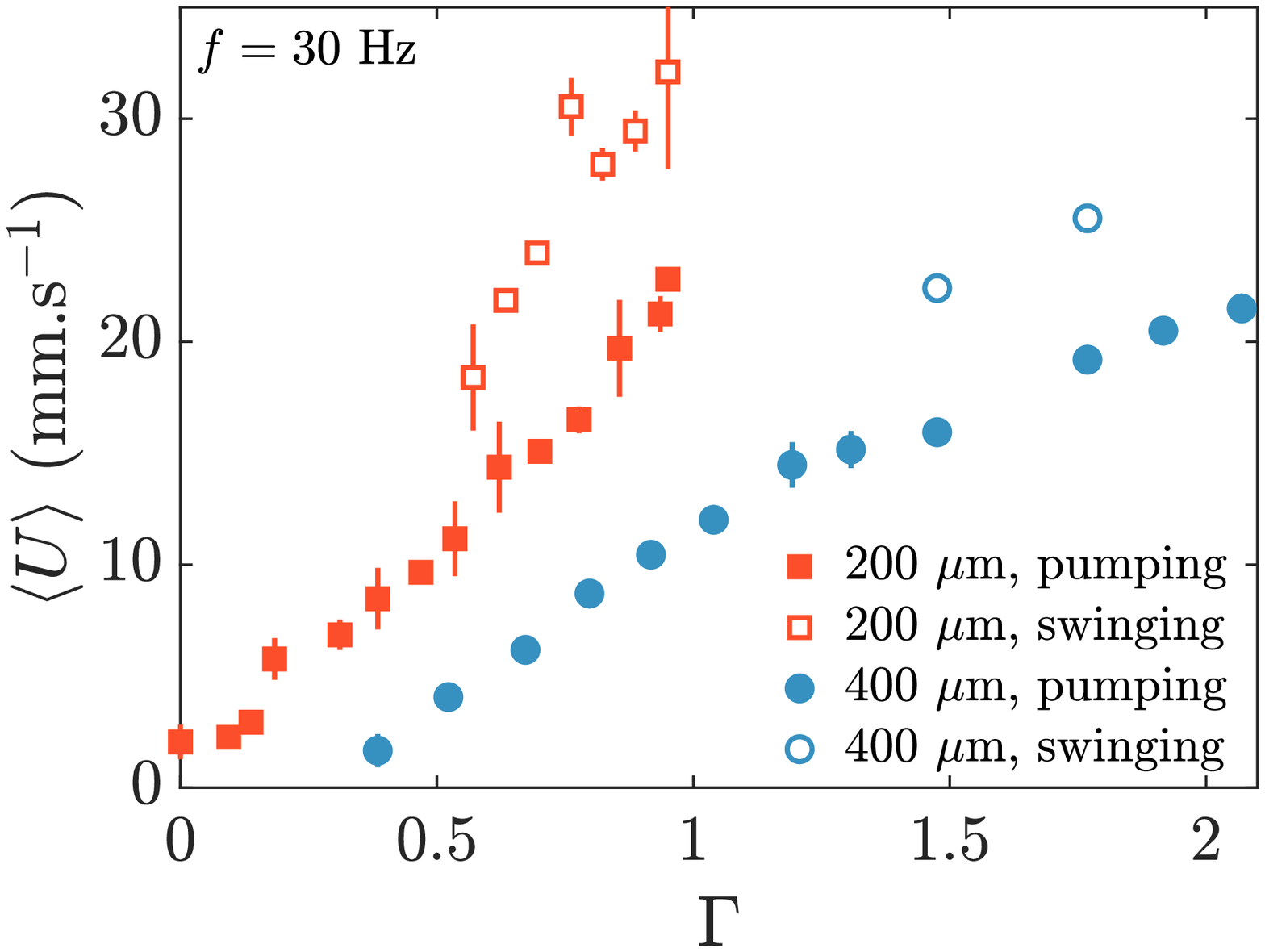}};
    	\node[] at (fig.north west) {$(a)$};
  	\end{tikzpicture}
  	\begin{tikzpicture}
    	\draw (0, 0) node[inner sep=0] (fig) {\includegraphics[width=0.47\textwidth]{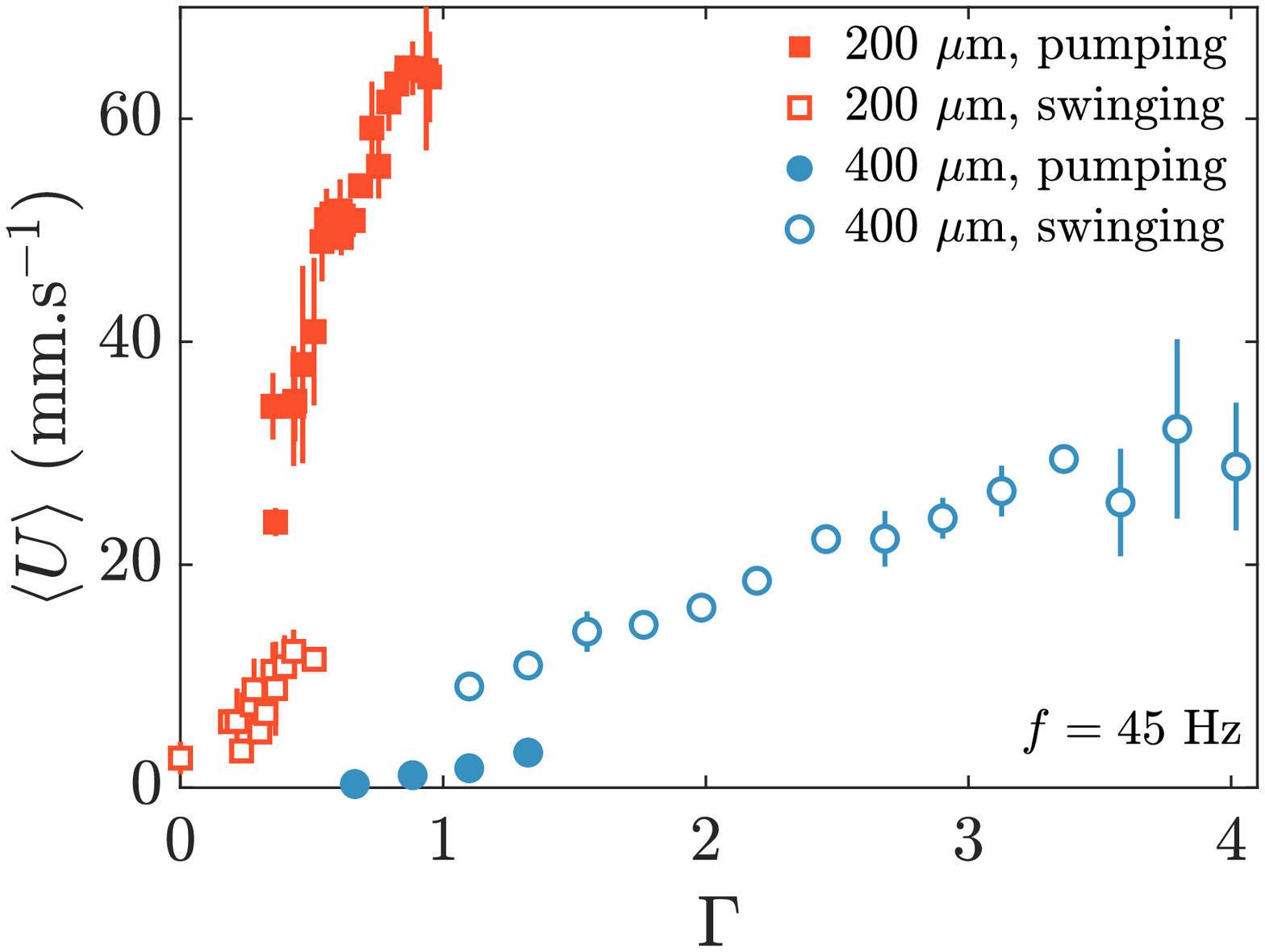}};
    	\node[] at (fig.north west) {$(b)$};
  	\end{tikzpicture}

  	\begin{tikzpicture}
    	\draw (0, 0) node[inner sep=0] (fig) {\includegraphics[width=0.47\textwidth]{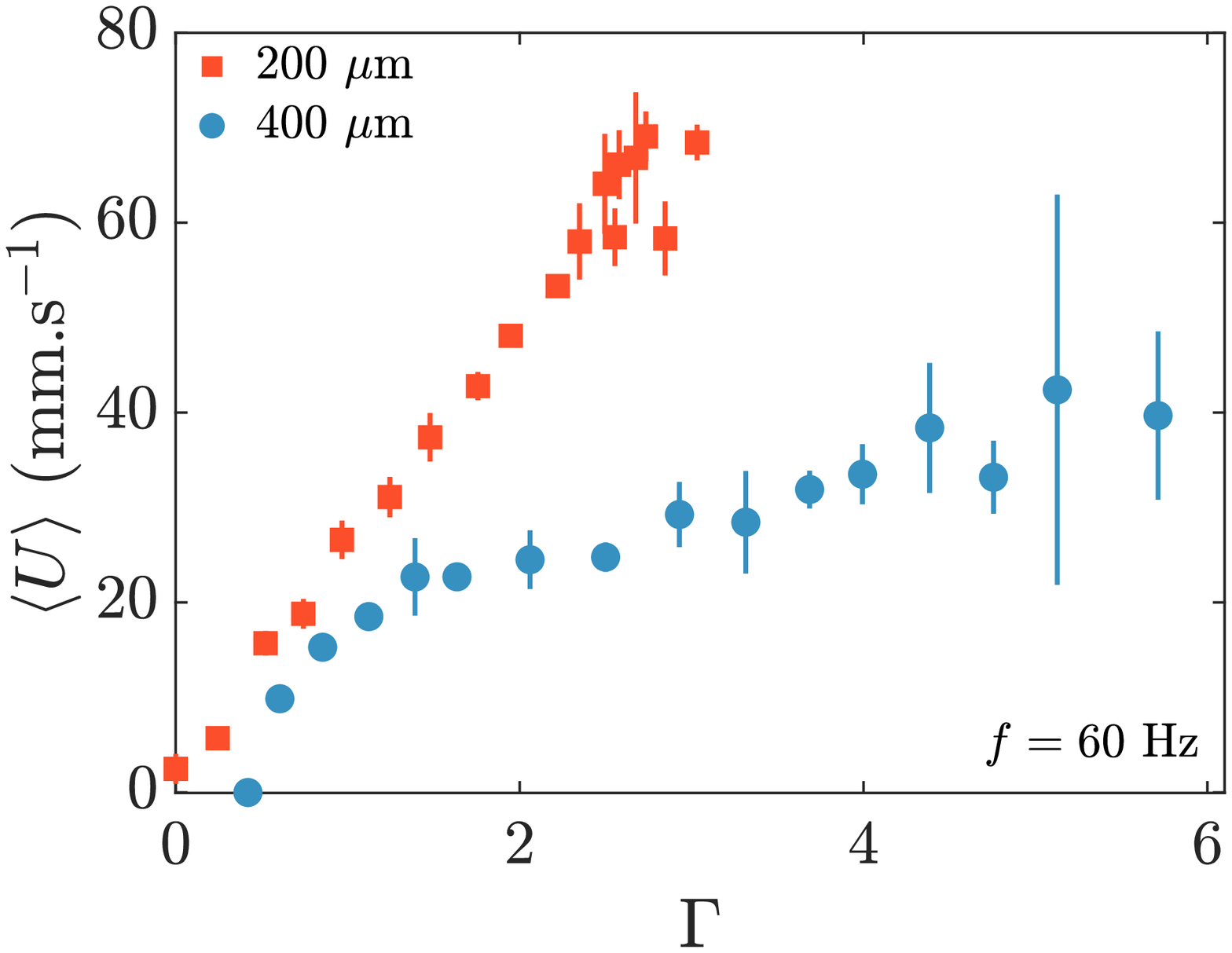}};
    	\node[] at (fig.north west) {$(c)$};
  	\end{tikzpicture}
  	\begin{tikzpicture}
    	\draw (0, 0) node[inner sep=0] (fig) {\includegraphics[width=0.47\textwidth]{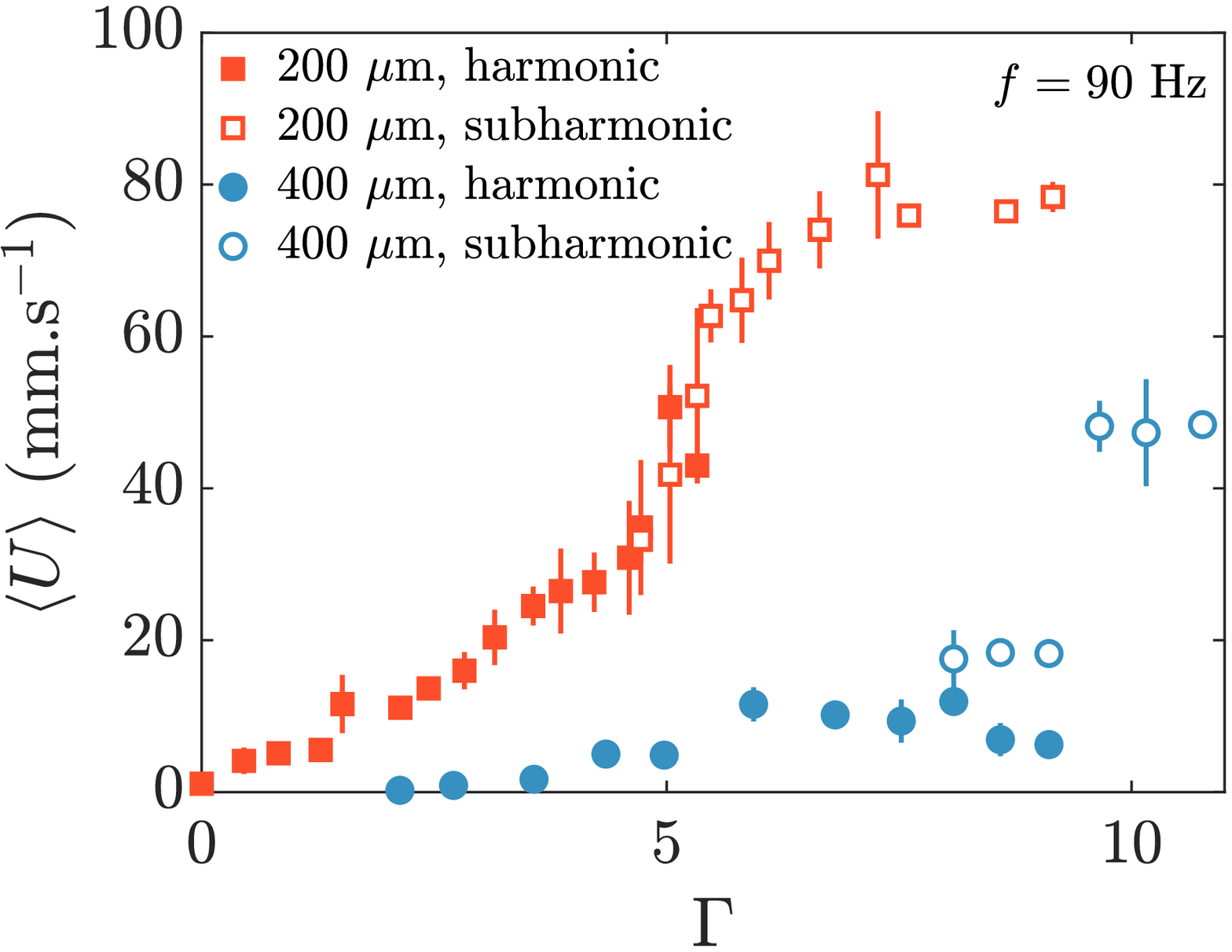}};
    	\node[] at (fig.north west){$(d)$};
  	\end{tikzpicture}
	\caption{
	Effect of the fibre diameter $b=200$ and 400 $\mu$m  on the sliding speed $\langle U \rangle$ as a function of the normalized acceleration. Here $V=4~\mu L$, $\alpha=27.5^\circ$.
	$(a,b)$ $f=30$ and 45 Hz, respectively. Filled symbols represent harmonic pumping, open symbols represent subharmonic swinging at $f/2$.
	$(c)$ $f=60$ Hz. Filled symbols represent harmonic pumping.
	$(d)$ $f=90$ Hz. Filled symbols represent harmonic pumping, open symbols represent subharmonic pumping at $f/2$.
	\label{fig:fibre_diameter}
	}
\end{figure}

\begin{figure}
	\centering
	\begin{tikzpicture}
    	\draw (0, 0) node[inner sep=0] (fig) {\includegraphics[width=0.47\textwidth]{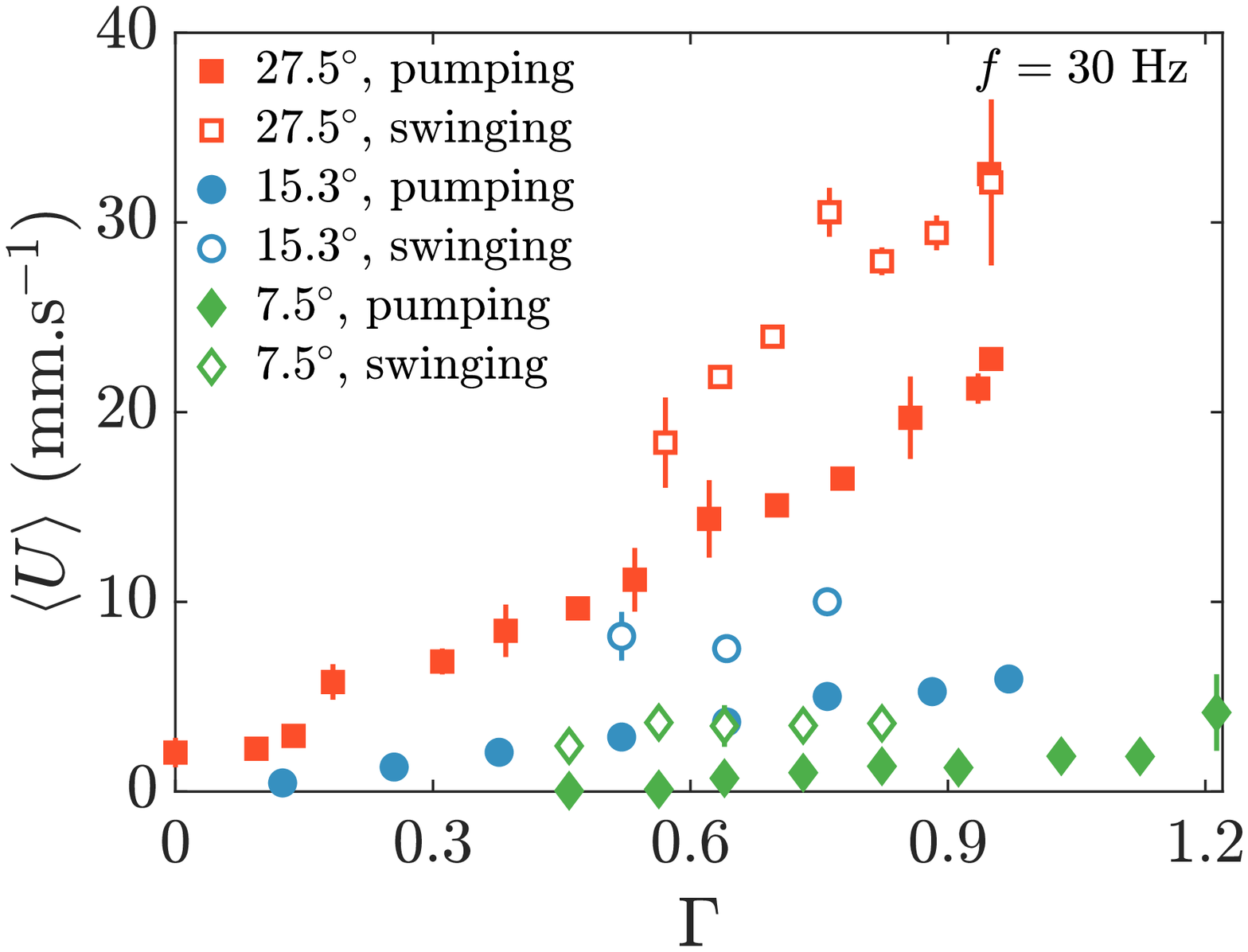}};
    	\node[] at (fig.north west) {$(a)$};
  	\end{tikzpicture}
  	\begin{tikzpicture}
    	\draw (0, 0) node[inner sep=0] (fig) {\includegraphics[width=0.47\textwidth]{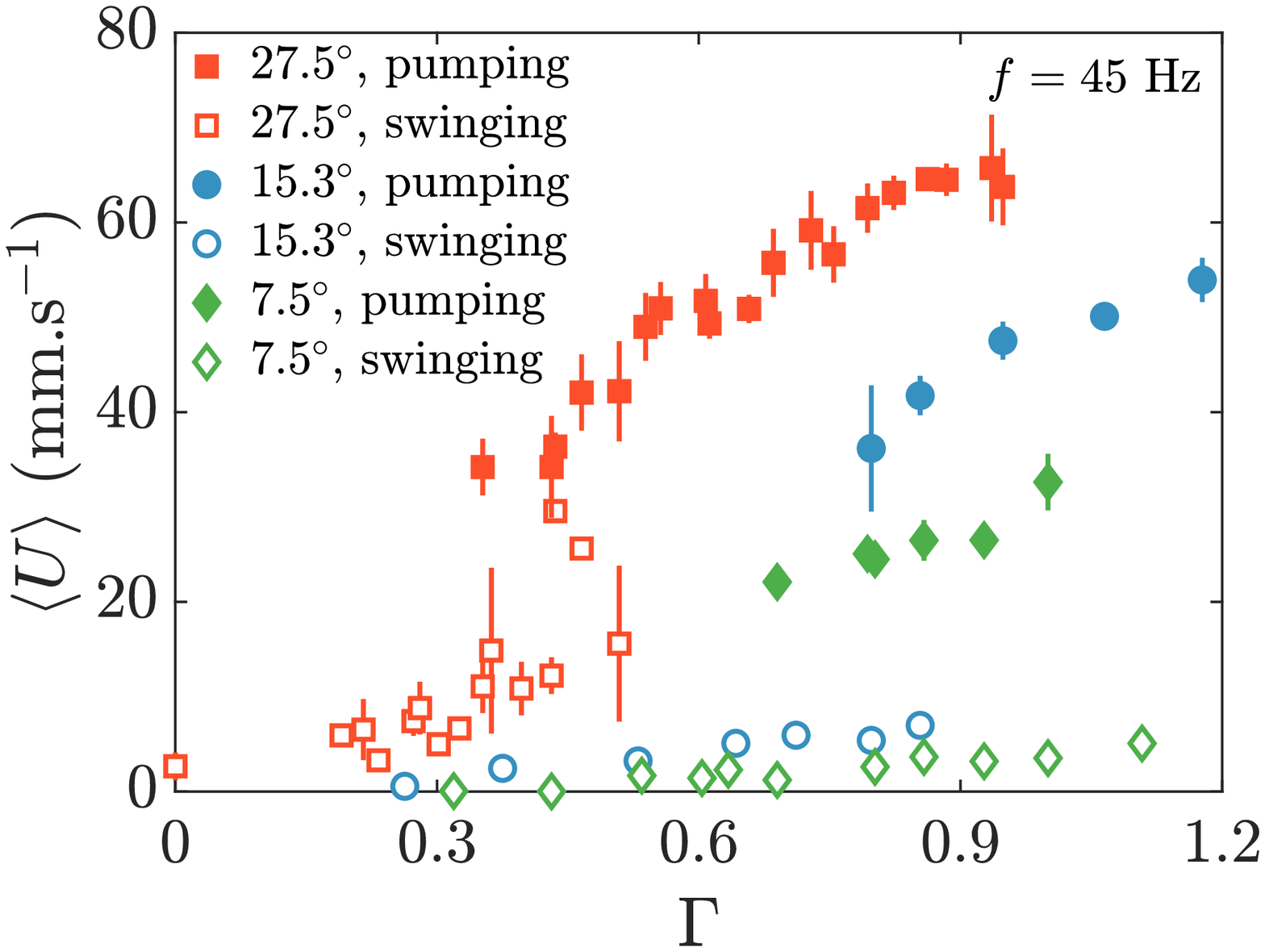}};
    	\node[] at (fig.north west) {$(b)$};
  	\end{tikzpicture}

  	\begin{tikzpicture}
    	\draw (0, 0) node[inner sep=0] (fig) {\includegraphics[width=0.47\textwidth]{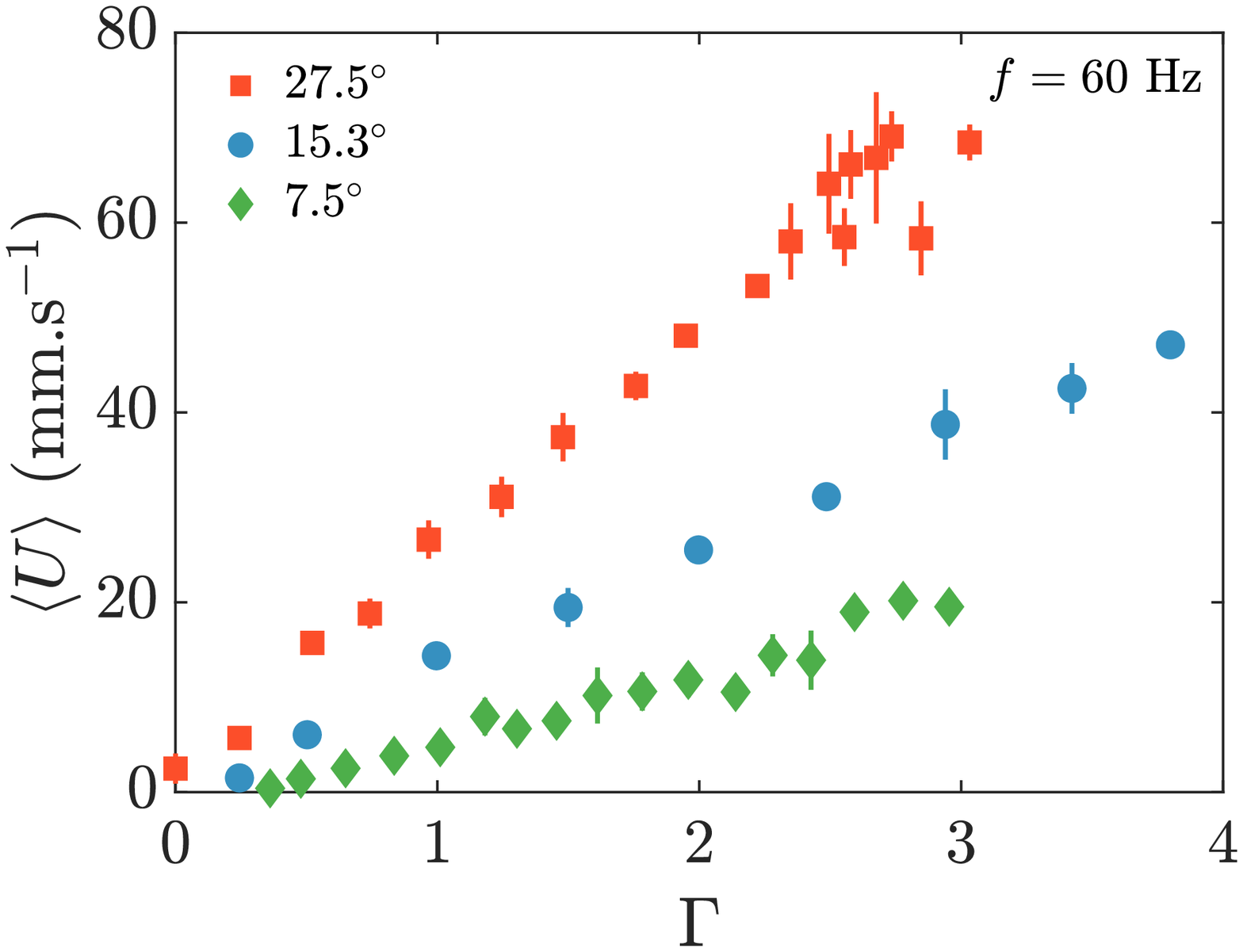}};
    	\node[] at (fig.north west) {$(c)$};
  	\end{tikzpicture}
  	\begin{tikzpicture}
    	\draw (0, 0) node[inner sep=0] (fig) {\includegraphics[width=0.47\textwidth]{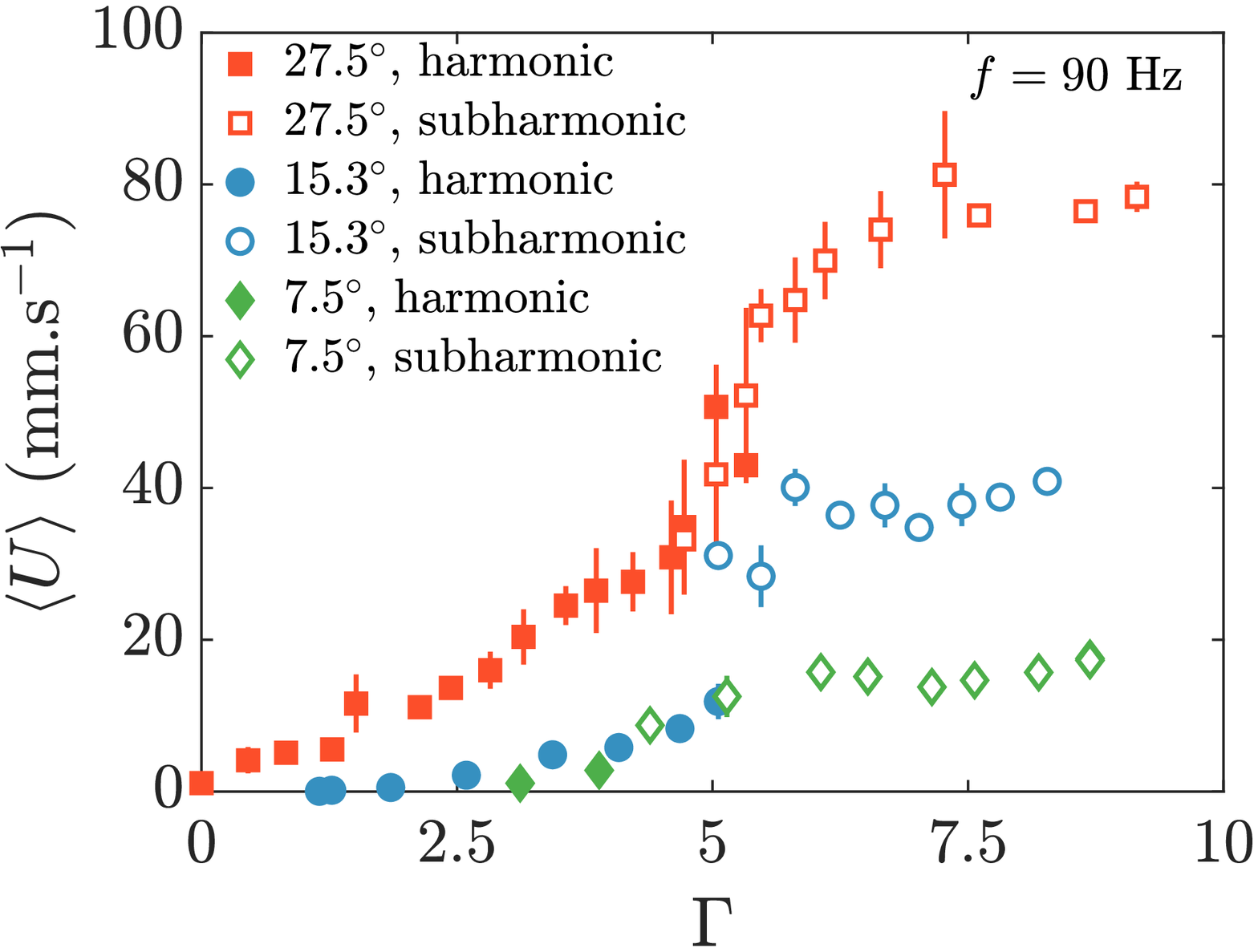}};
    	\node[] at (fig.north west){$(d)$};
  	\end{tikzpicture}
	\caption{
	Effect of the angle  $\alpha=27.5,~15.3$ and $7.5^\circ$  on the sliding speed $\langle U \rangle$ as a function of the normalized acceleration. Here $V=4~\mu L$, $b=200~\mu$m.
	$(a,b)$ $f=30$ and 45 Hz, respectively. Filled symbols represent harmonic pumping, open symbols represent subharmonic swinging at $f/2$.
	$(c)$ $f=60$ Hz. Filled symbols represent harmonic pumping.
	$(d)$ $f=90$ Hz. Filled symbols represent harmonic pumping, open symbols represent subharmonic pumping at $f/2$.
	\label{fig:fibre_angle}
	}
\end{figure}

Here we present additional data regarding the speed $\langle U \rangle$ and mode of response of the droplet as a function of the amplitude of oscillations for 4 frequencies: $f=30$, 45, 60 and 90 Hz.
The parameters are variations for those used in figure \ref{fig:speed_frequency} ($b=200~\mu$m, $V=4~\mu$L, $\alpha=27.5^\circ$):
we changed the fibre diameter $b$ to 400 $\mu$m in figure \ref{fig:fibre_diameter}, and decreased the tilt angle $\alpha$ down to $15^\circ$ and $7.5 ^\circ$ in figure \ref{fig:fibre_angle}.
In both cases this leads to droplets being pinned on the fibre without oscillations: $U_0=0$.

\section{Governing equations of an elastic pendulum}
\label{appendix:elasticpendulum}

We consider a pendulum in a two dimensional plane consisting of a point mass $m$ attached to a support by a linear spring of constant $k$ and length $L=L_0+r$, with $L_0$ the length of the spring at rest and $r$ its extension, see figure \ref{fig:pendulum}$(c)$.
The pendulum makes an angle $\phi$ with respect to the vertical ; it is placed in a gravitational field $g$ and its support oscillates vertically as $A\cos(\omega t)$. We let $\Gamma=A\omega^2/g$ the normalized maximal acceleration due to these oscillations.
The dynamics of the system is governed by the following nonlinear system of ordinary differential equations \citep{NayfehBook}:
\begin{subequations}
\begin{align}
    r'' + 2\alpha_1 r' + \frac{k}{m}r - L {\phi}'^2 =  g(1-\Gamma \cos\omega t)\cos\phi,\\
    \phi'' + 2\alpha_2 \phi' + \frac{g}{L}\left(1-\Gamma \cos \omega t\right) \sin \phi + \frac 2L r' {\phi}'=0.
\end{align}
    \label{eq:elasticpendulumdimension}
\end{subequations}
Here $(.)'$ represent the derivative with respect to the time $t$, $\alpha_1$ and $\alpha_2$ are two positive coefficients modelling linear damping.
At $\phi=0$ and without forcing ($\Gamma=0$) the pendulum has an equilibrium length $L_{\rm eq}=L_0 + gm/k$, and we let $\rho=L_0+r-L_{\rm eq}$ the extension with respect to $L_{\rm eq}$.
We also introduce the following variables in order to make \eqref{eq:elasticpendulumdimension} dimensionless:
\begin{equation}
\begin{aligned}
    \omega_s=\left(\frac km\right)^{1/2}, \quad \omega_p=\left(\frac{g}{L_{\rm eq}}\right)^{1/2},\quad \Omega=\frac{\omega_p}{\omega_s}, \quad 
    \eta=\frac{\omega}{\omega_p}, \\
    c_s=\frac{\alpha_s}{\omega_s}, \quad c_p=\frac{\alpha_p}{\omega_p},R=\frac{\rho}{L_{\rm eq}}, \quad \tau=\omega t,
\end{aligned}
\end{equation}
with $\omega_s$ and $\omega_p$ the natural frequencies of oscillations of the spring and of the pendulum, respectively.
We use $\dot{(.)}$ to denote derivatives with respect to the dimensionless time $\tau$, so that the dimensionless version of \eqref{eq:elasticpendulumdimension} is
\begin{subequations}
\begin{align}
    \ddot{R} + 2\frac{c_s}{\eta\Omega}\dot{R} + \frac{1}{\eta^2\Omega^2}R - \left(1+R\right)\dot\phi^2  &=  -\frac{1}{\eta^2}\left(1-\cos\phi+\Gamma\cos\tau\right),\\
    \ddot{\phi} + 2\left(\frac{c_p}{\eta}+\frac{\dot R}{1+R}\right) \dot{\phi} + \frac{1}{\eta^2} \frac{1-\Gamma\cos\tau}{1+R}\sin\phi &= 0.
\end{align}
    \label{eq:elasticpendulumdimensionless}
\end{subequations}
This system is governed by 5 dimensionless parameters: $\Omega$, $\eta$, $\Gamma$, $c_s$ and $c_p$.
We now assume small angles $\phi=\mathcal O(\epsilon)$ with $\lvert\epsilon\rvert\ll 1$. Neglecting terms of order $\epsilon^2$ and higher, \eqref{eq:elasticpendulumdimensionless} simplifies to:
\begin{subequations}
\begin{align}
    \ddot{R} + 2\frac{c_s}{\eta\Omega}\dot{R} + \frac{1}{\eta^2\Omega^2}R  &=  -\frac{\Gamma}{\eta^2}\cos\tau,
    \label{eq:equationspring}\\
    \ddot{\phi} + 2\left(\frac{c_p}{\eta}+\frac{\dot R}{1+R}\right) \dot{\phi} + \frac{1}{\eta^2} \frac{1-\Gamma\cos\tau}{1+R}\phi &= 0.
    \label{eq:pendulum}
\end{align}
    \label{eq:elasticpendulumlinear}
\end{subequations}
Under this limit of small angles, \eqref{eq:elasticpendulumlinear} exhibits one-way coupling: the pendulum mode governed by $\phi$ depends on the spring mode governed by $R$, but $R$ is independent of $\phi$.

We first assume small oscillations of the spring with $R=\mathcal{O}(\epsilon)$. The equation for the pendulum \eqref{eq:pendulum} is then independent of $R$ and is given by the following damped Mathieu equation
\begin{align}
    \ddot{\phi}+2\frac{c_p}{\eta}\dot{\phi}+\frac{1}{\eta^2}\left(1-\Gamma\cos\tau\right)\phi=0.
    \label{eq:mathieu_adim}
\end{align}
The dimensional version of \eqref{eq:mathieu_adim} is \eqref{eq:Mathieu}, discussed in \S\ref{subsec:forcedpendulum}.

Assuming now $R=\mathcal{O}(1)$, the variations of $R$ cannot be neglected. The solution of \eqref{eq:equationspring} consists of a transient regime that exponentially decays and that we neglect, and of  harmonic oscillations given by
\begin{align}
    R(\tau)=\frac{-\Omega^2}{\left(1-\eta^2\Omega^2\right)^2 + 4c_s^2\eta^2\Omega^2}\left[\Gamma\left(1-\eta^2\Omega^2\right)\cos\tau \right.% \\
    \left. + 2c_s\eta\Omega\sin\tau \right],
    \label{eq:spring}
\end{align}
while $\phi$ is still governed by \eqref{eq:pendulum}.
The dimensional version of this system is \eqref{eq:forcedMathieuModified} discussed in \S \ref{subsec:forcedelasticpendulum}.

\begin{figure}
	\centering
	\begin{tikzpicture}
    	\draw (0, 0) node[inner sep=0] (fig) {\includegraphics[width=0.32\textwidth]{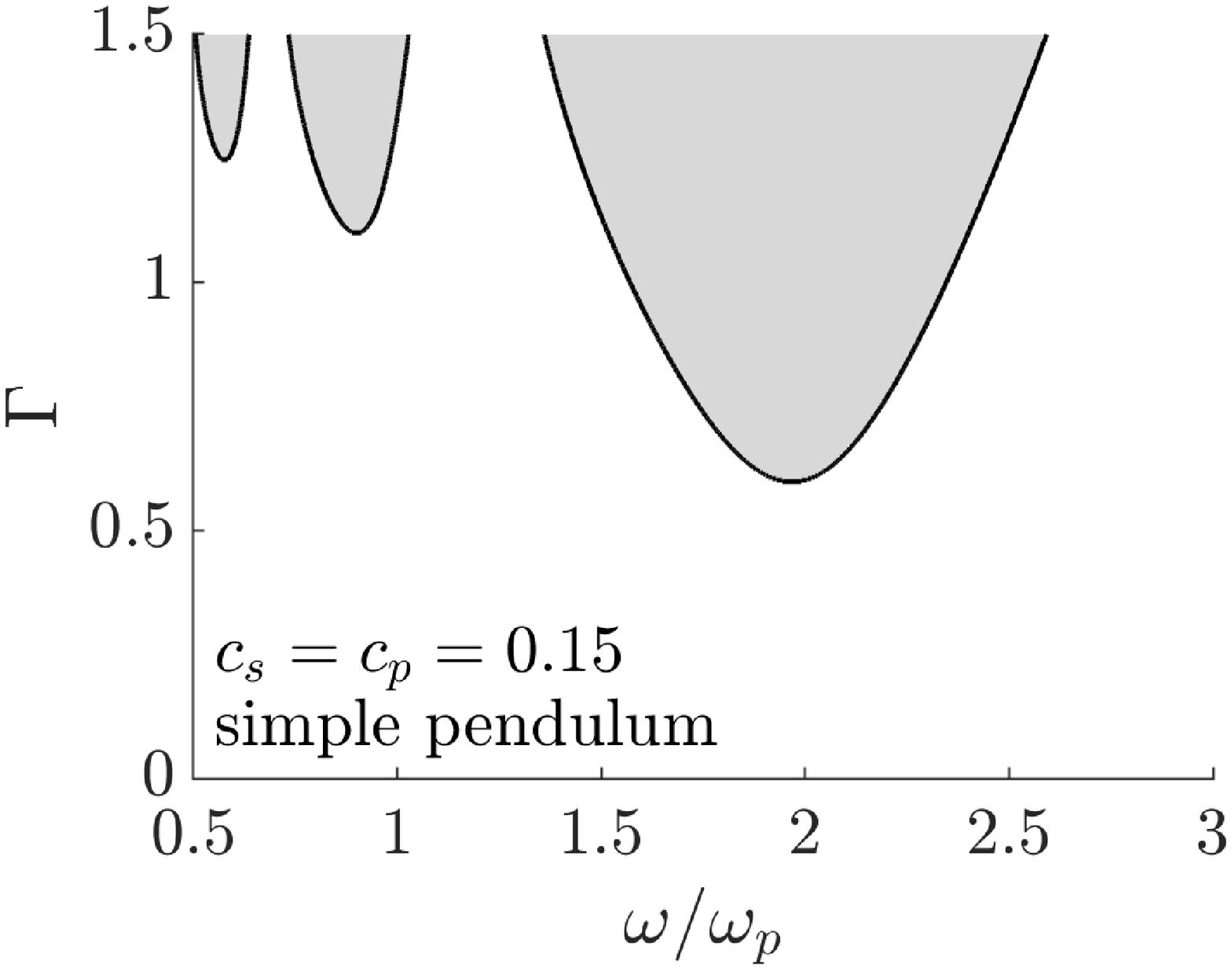}};
    	%\node[] at (fig.north west) {$(a)$};
  	\end{tikzpicture}
  	\begin{tikzpicture}
    	\draw (0, 0) node[inner sep=0] (fig) {\includegraphics[width=0.32\textwidth]{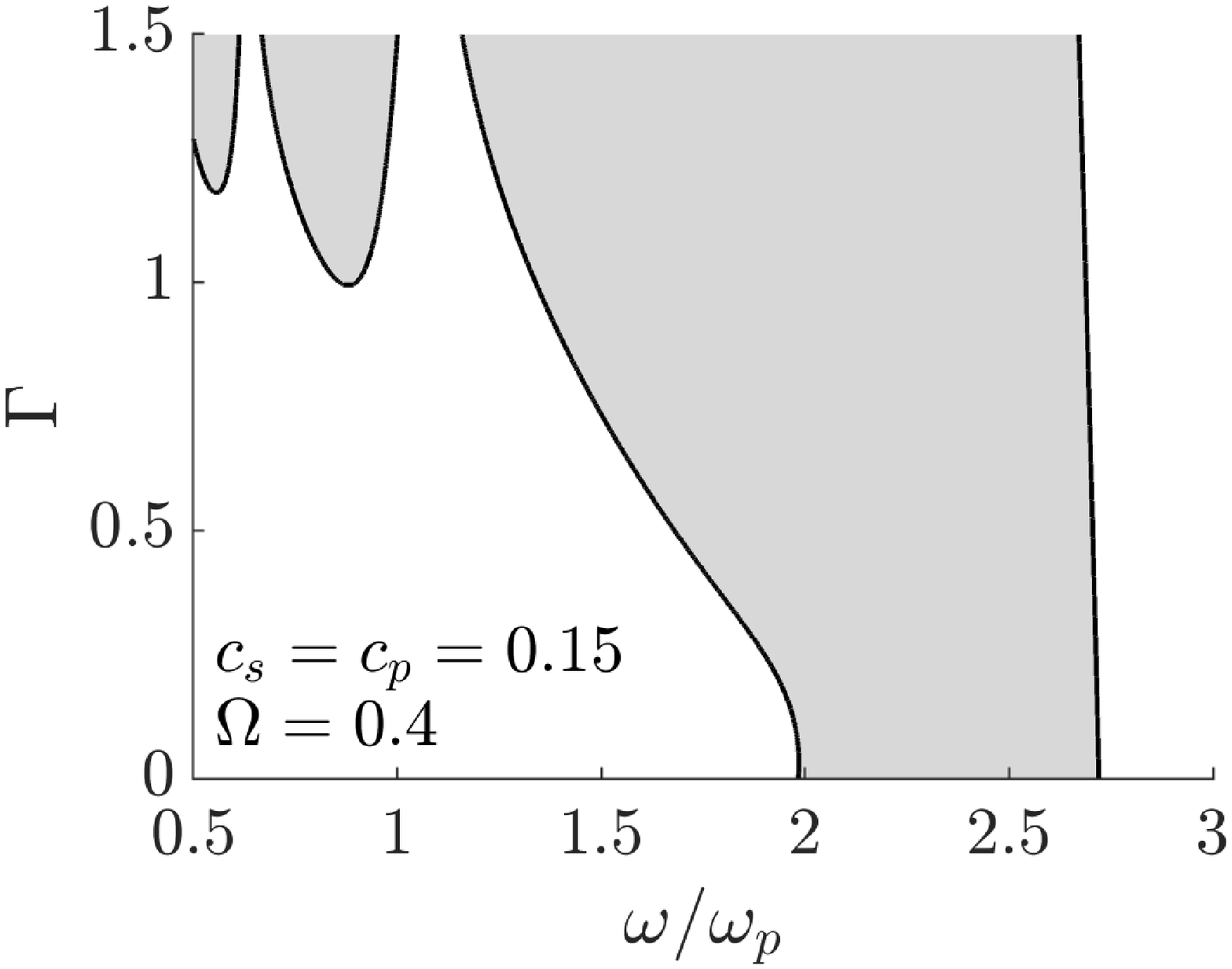}};
    	%\node[] at (fig.north west) {$(b)$};
  	\end{tikzpicture}
  	\begin{tikzpicture}
    	\draw (0, 0) node[inner sep=0] (fig) {\includegraphics[width=0.32\textwidth]{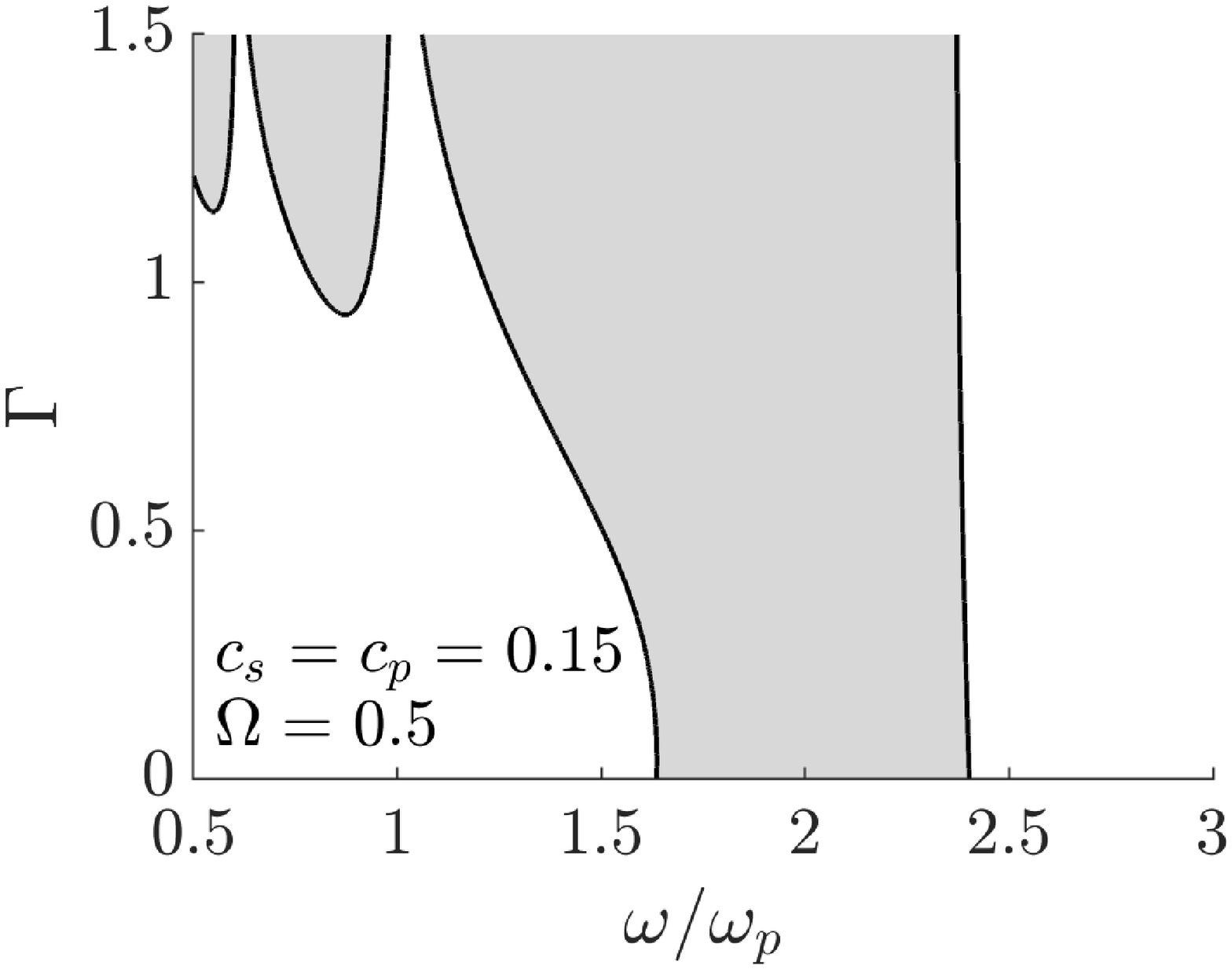}};
    	%\node[] at (fig.north west) {$(c)$};
  	\end{tikzpicture}

  	\begin{tikzpicture}
    	\draw (0, 0) node[inner sep=0] (fig) {\includegraphics[width=0.32\textwidth]{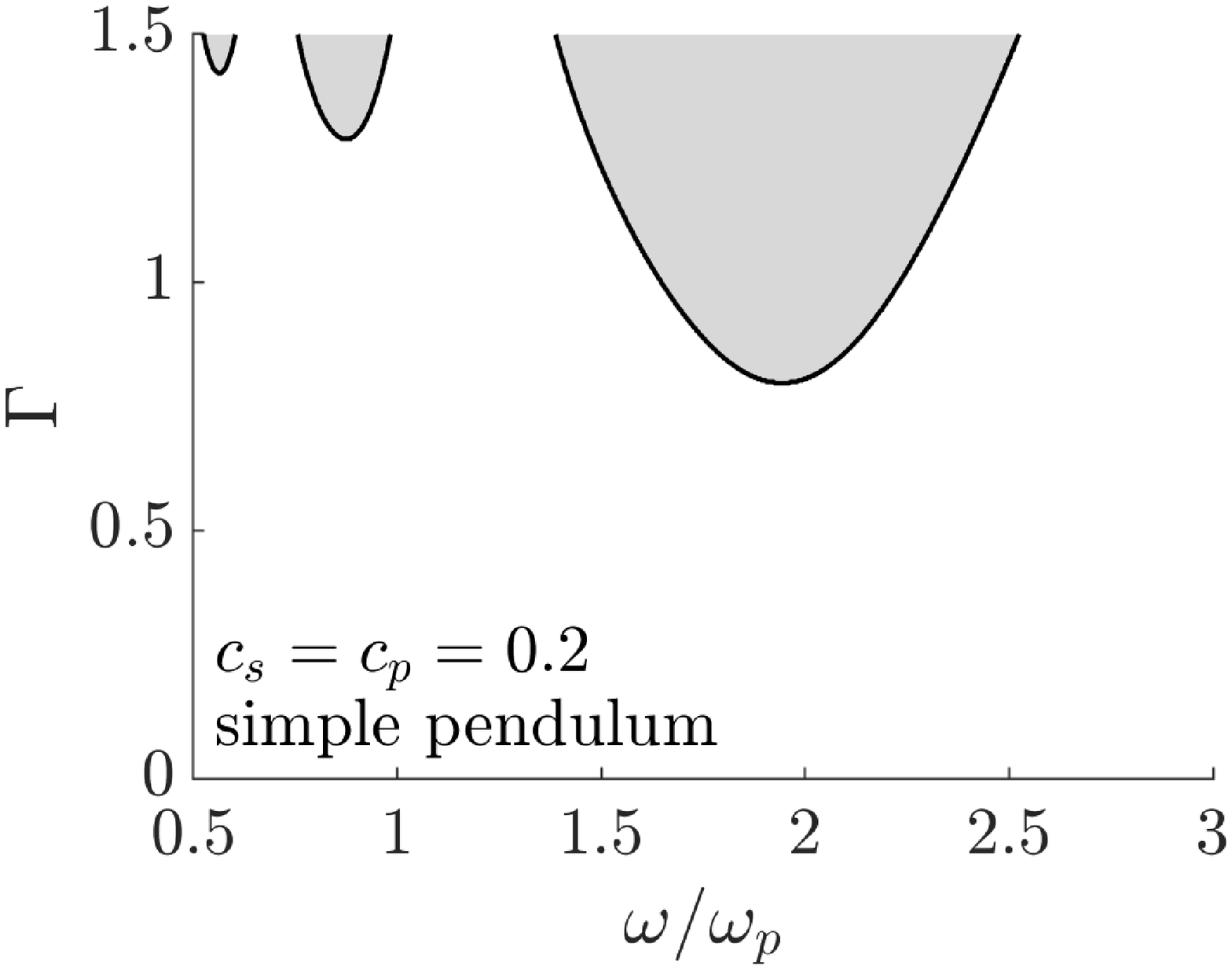}};
    	%\node[] at (fig.north west) {$(a)$};
  	\end{tikzpicture}
  	\begin{tikzpicture}
    	\draw (0, 0) node[inner sep=0] (fig) {\includegraphics[width=0.32\textwidth]{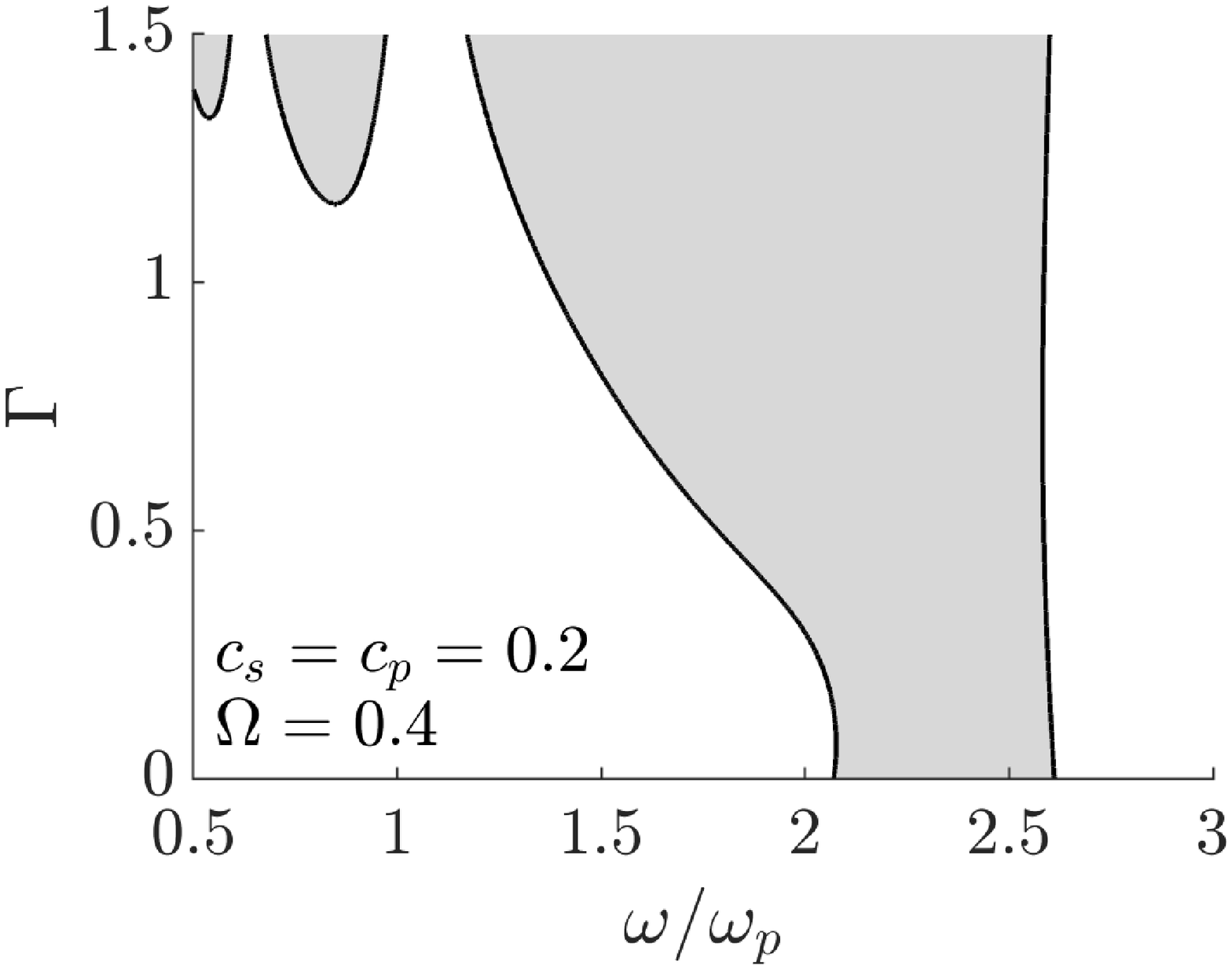}};
    	%\node[] at (fig.north west) {$(b)$};
  	\end{tikzpicture}
  	\begin{tikzpicture}
    	\draw (0, 0) node[inner sep=0] (fig) {\includegraphics[width=0.32\textwidth]{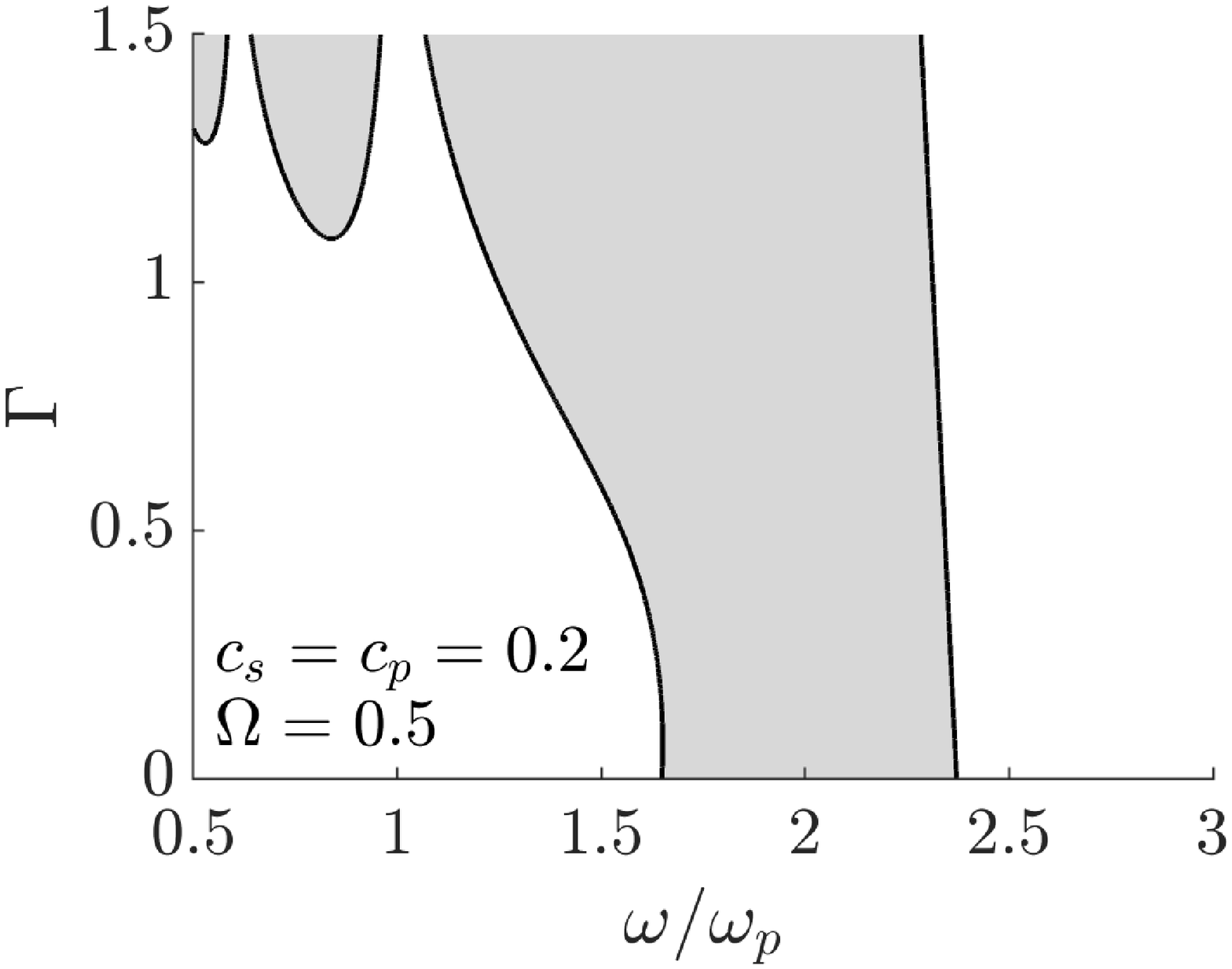}};
    	%\node[] at (fig.north west) {$(c)$};
  	\end{tikzpicture}

  	\begin{tikzpicture}
    	\draw (0, 0) node[inner sep=0] (fig) {\includegraphics[width=0.32\textwidth]{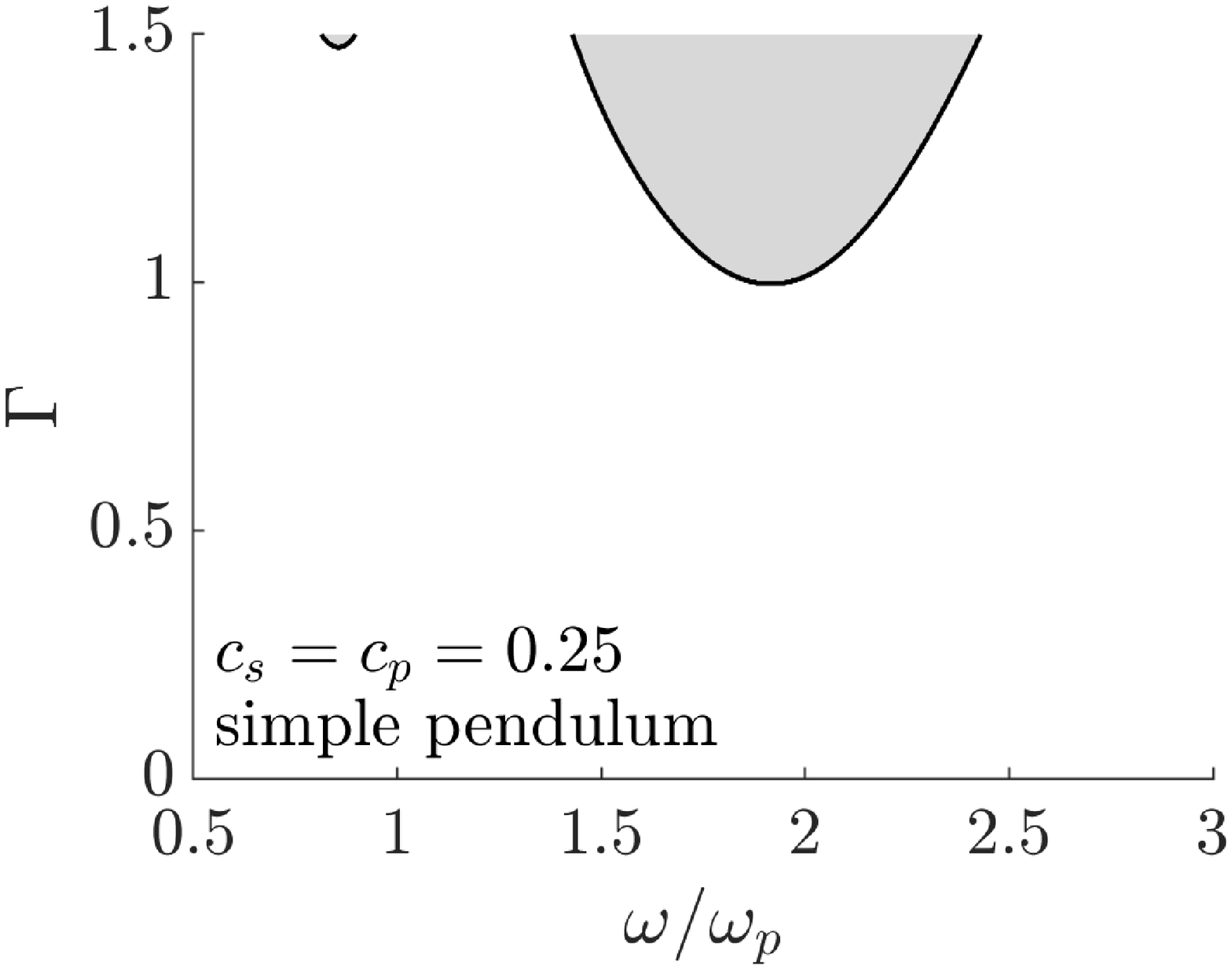}};
    	%\node[] at (fig.north west) {$(a)$};
  	\end{tikzpicture}
  	\begin{tikzpicture}
    	\draw (0, 0) node[inner sep=0] (fig) {\includegraphics[width=0.32\textwidth]{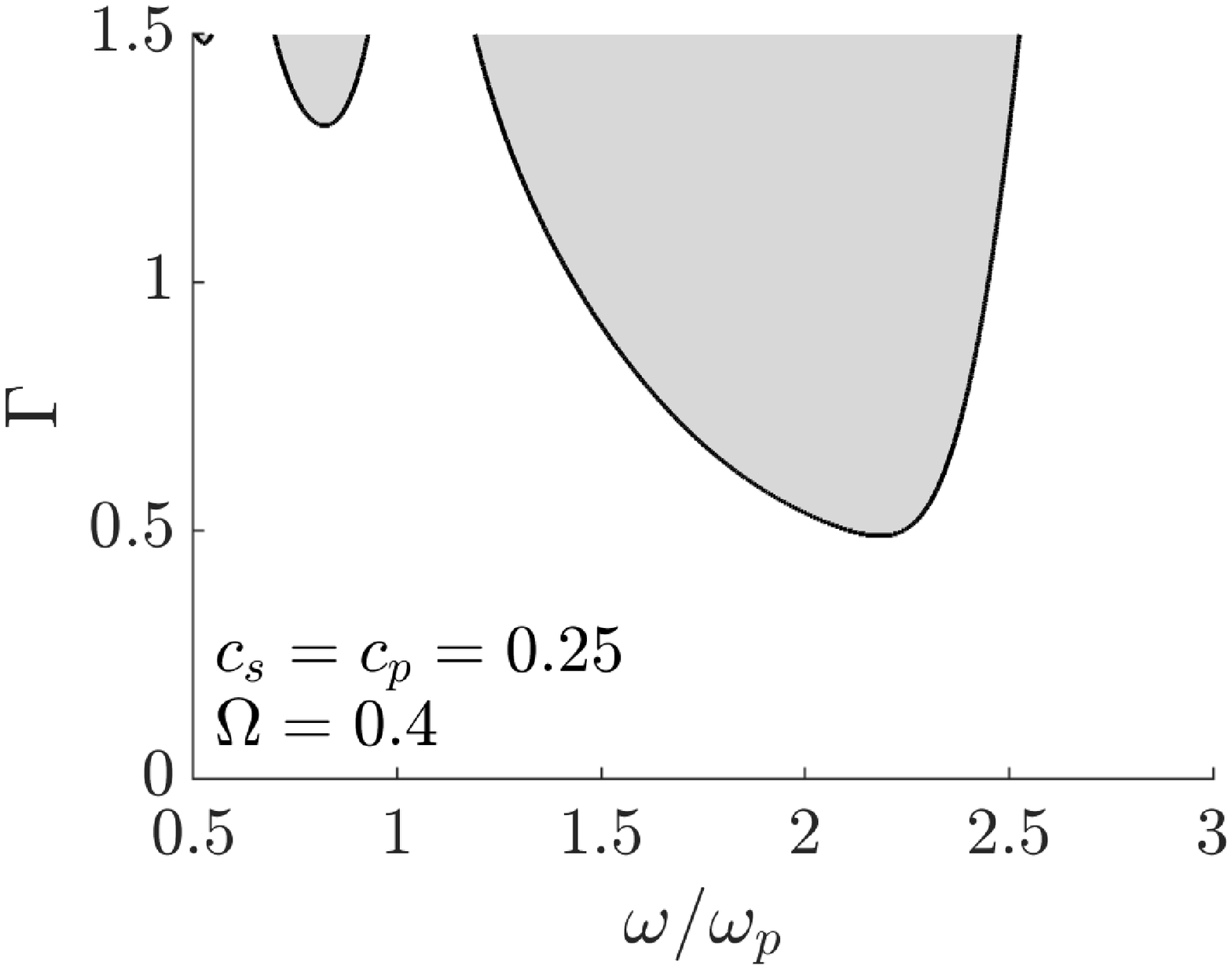}};
    	%\node[] at (fig.north west) {$(b)$};
  	\end{tikzpicture}
  	\begin{tikzpicture}
    	\draw (0, 0) node[inner sep=0] (fig) {\includegraphics[width=0.32\textwidth]{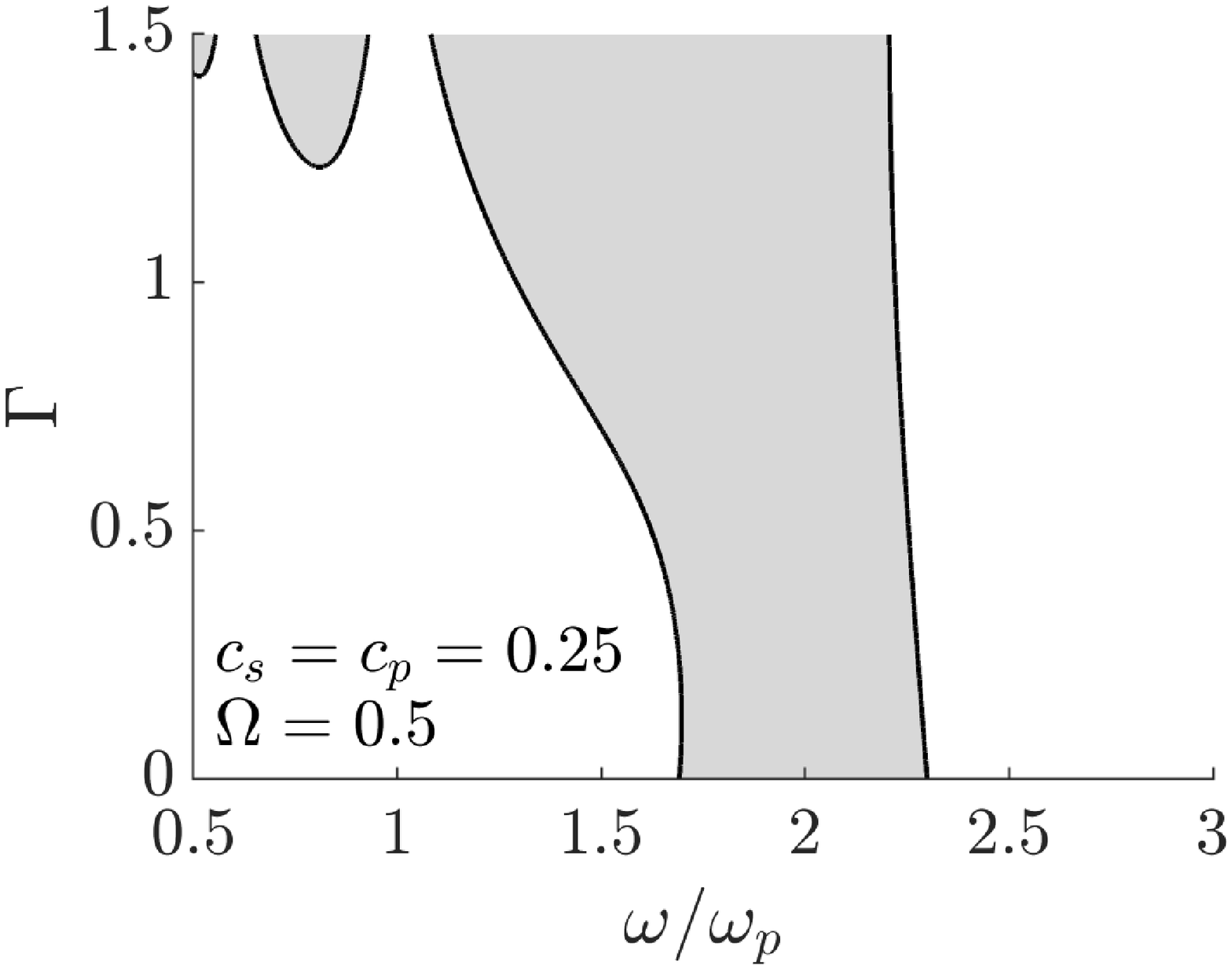}};
    	%\node[] at (fig.north west) {$(c)$};
  	\end{tikzpicture}

  	\begin{tikzpicture}
    	\draw (0, 0) node[inner sep=0] (fig) {\includegraphics[width=0.32\textwidth]{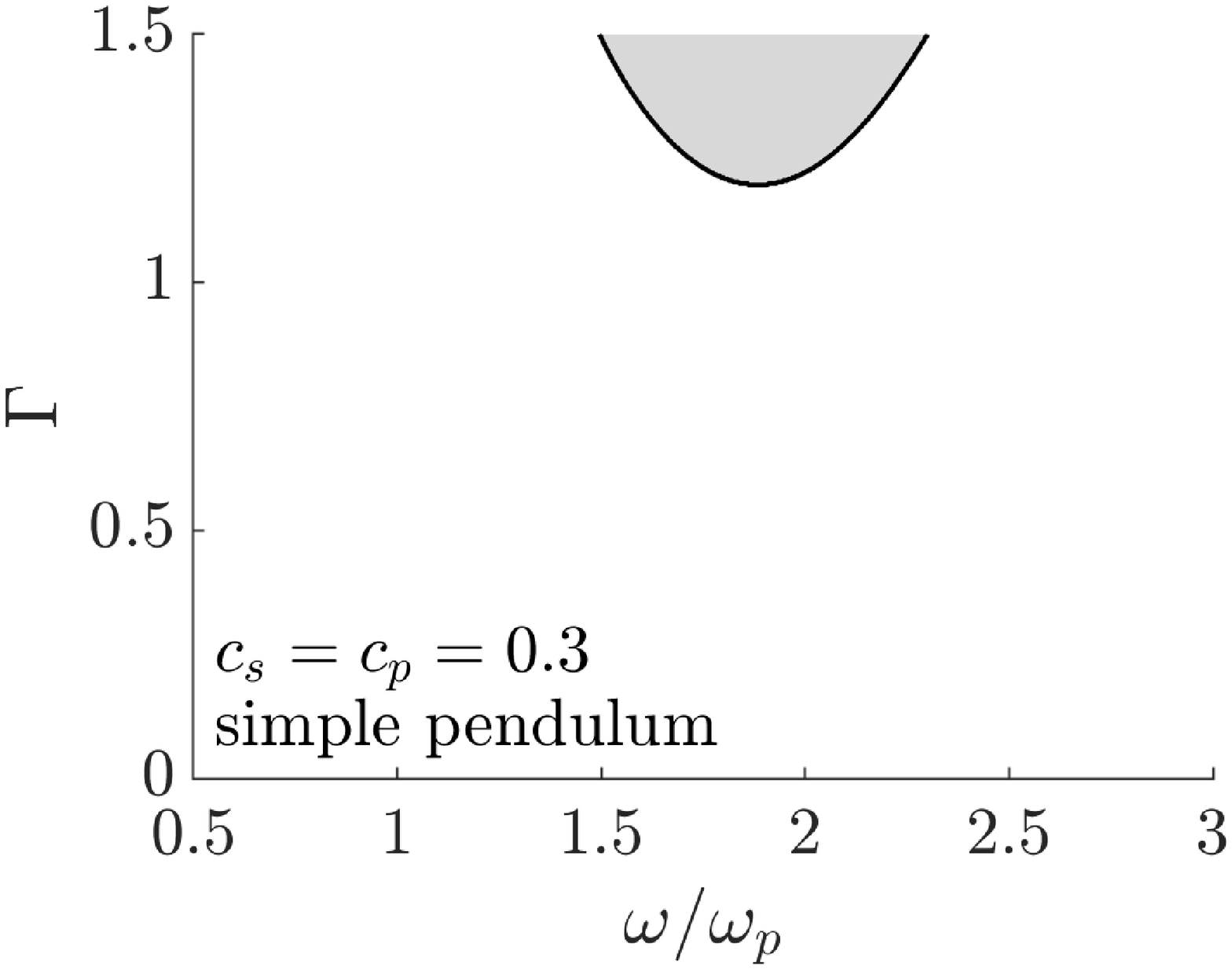}};
    	%\node[] at (fig.north west) {$(a)$};
  	\end{tikzpicture}
  	\begin{tikzpicture}
    	\draw (0, 0) node[inner sep=0] (fig) {\includegraphics[width=0.32\textwidth]{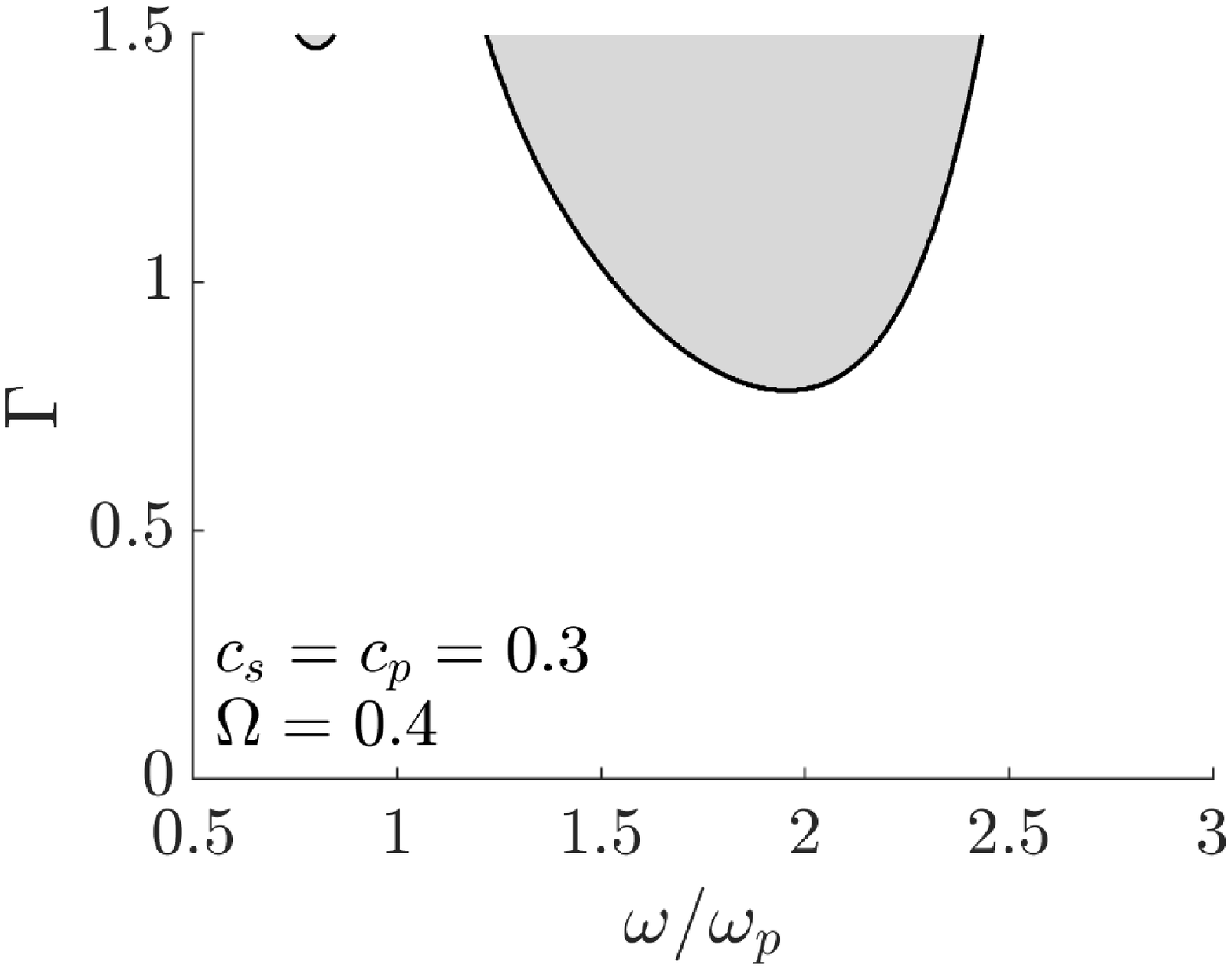}};
    	%\node[] at (fig.north west) {$(b)$};
  	\end{tikzpicture}
  	\begin{tikzpicture}
    	\draw (0, 0) node[inner sep=0] (fig) {\includegraphics[width=0.32\textwidth]{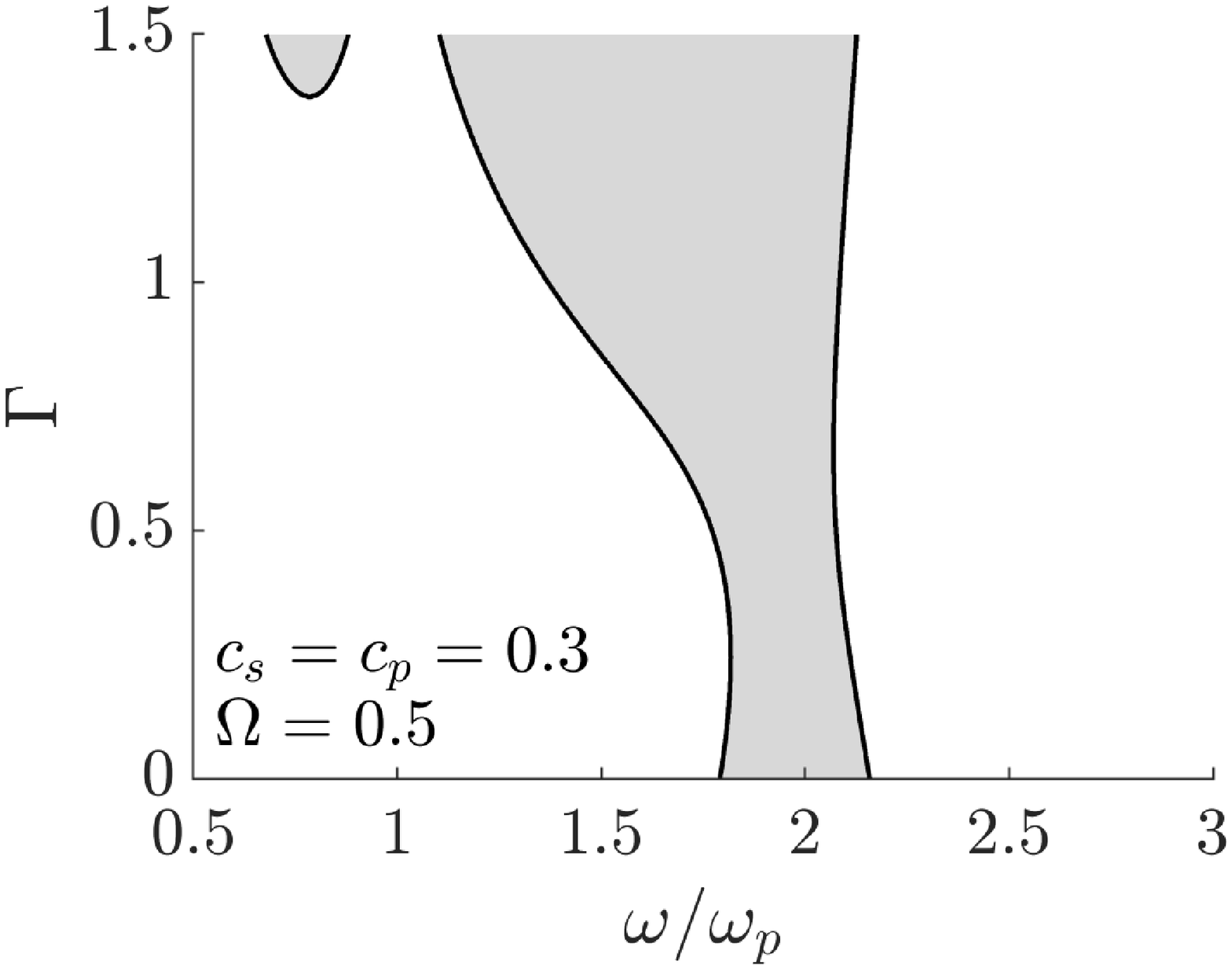}};
    	%\node[] at (fig.north west) {$(c)$};
  	\end{tikzpicture}
	\caption{
	Linear stability diagrams for (left column) a simple pendulum modeled by \eqref{eq:mathieu_adim} (\eqref{eq:Mathieu} in the main text) or an elastic pendulum modeled by \eqref{eq:pendulum} and \eqref{eq:spring}
(\eqref{eq:forcedMathieuModified} in the main text) with (middle column) $\Omega=0.4$ and (right column) $\Omega=0.5$.
	Different rows show different damping coefficients $c=c_p=c_s$ going from (top row) $c=0.15$ and up to 0.2, 0.25 and (bottom row) $c=0.3$.
	Shaded areas represent regions where the position $\phi=0$ is unstable.
	\label{fig:elasticpend_c}
	}
\end{figure}

The linear stability diagrams of the position $\phi=0$ of \eqref{eq:elasticpendulumlinear}, or \eqref{eq:Mathieu} and \eqref{eq:forcedMathieuModified} in \S \ref{sec:swinging}, are shown in figure \ref{fig:pendulum}$(d)$ with $c_p=c_s=0.15$. Additional diagrams for larger damping coefficients are shown in figure \ref{fig:elasticpend_c}. 
They are obtained using Floquet theory, the theory of linear ordinary differential equations with time-periodic coefficients, combined with numerical integration  \citep{Cesari2012,Kovacic2018}.
We note that we have only considered damping coefficients such that $c_s\geq0.15$. For smaller $c_s$  \eqref{eq:spring} shows that the length of the pendulum $1+R$ could become negative. Considering an important damping of the spring motion therefore allows us to study this system without adding nonlinearities which would further complicate the model and analysis.

%\begin{comment}
%\end{comment}

\bibliographystyle{jfm}
\bibliography{./references}

\begin{thebibliography}{41}
\expandafter\ifx\csname natexlab\endcsname\relax\def\natexlab#1{#1}\fi
\def\au#1{#1} \def\ed#1{#1} \def\yr#1{#1}\def\at#1{#1}\def\jt#1{\textit{#1}}
  \def\bt#1{#1}\def\bvol#1{\textbf{#1}} \def\vol#1{#1} \def\pg#1{#1}
  \def\publ#1{#1}\def\arxiv#1{#1}\def\org#1{#1}\def\st#1{\textit{#1}}

\bibitem[Agranovski \& Braddock(1998)]{Agranovski1998}
{\sc \au{Agranovski, I.E.} \& \au{Braddock, R.D.}} \yr{1998}  \at{Filtration of
  liquid aerosols on wettable fibrous filters}.  \jt{AIChE journal}
  \bvol{44}~(12),  \pg{2775--2783}.

\bibitem[Bick {\em et~al.\/}(2015)Bick, Boulogne, Sauret \& Stone]{Bick2015}
{\sc \au{Bick, A.}, \au{Boulogne, F.}, \au{Sauret, A.} \& \au{Stone, H.A.}}
  \yr{2015}  \at{Tunable transport of drops on a vibrating inclined fiber}.
  \jt{Appl. Phys. Lett.}  \bvol{107}~(18),  \pg{181604}.

\bibitem[Bintein {\em et~al.\/}(2019)Bintein, Bense, Clanet \&
  Qu{\'e}r{\'e}]{Bintein2019}
{\sc \au{Bintein, P.-B.}, \au{Bense, H.}, \au{Clanet, C.} \& \au{Qu{\'e}r{\'e},
  D.}} \yr{2019}  \at{Self-propelling droplets on fibres subject to a
  crosswind}.  \jt{Nat. Phys.}  \bvol{15}~(10),  \pg{1027--1032}.

\bibitem[Bostwick \& Steen(2014)]{Bostwick2015}
{\sc \au{Bostwick, J.B.} \& \au{Steen, P.H.}} \yr{2014}  \at{Dynamics of
  sessile drops. {P}art 1. {I}nviscid theory}.  \jt{J. Fluid Mech.}
  \bvol{760},  \pg{5--38}.

\bibitem[Bradshaw \& Billingham(2018)]{Bradshaw2018}
{\sc \au{Bradshaw, J.T.} \& \au{Billingham, J.}} \yr{2018}  \at{Thick drops
  climbing uphill on an oscillating substrate}.  \jt{J. Fluid Mech.}
  \bvol{840},  \pg{131--153}.

\bibitem[Brunet {\em et~al.\/}(2007)Brunet, Eggers \& Deegan]{Brunet2007}
{\sc \au{Brunet, P.}, \au{Eggers, J.} \& \au{Deegan, R.D.}} \yr{2007}
  \at{Vibration-induced climbing of drops}.  \jt{Phys. Rev. Lett.}
  \bvol{99}~(14),  \pg{144501}.

\bibitem[Cesari(1971)]{Cesari2012}
{\sc \au{Cesari, L.}} \yr{1971} {\em Asymptotic behavior and stability problems
  in ordinary differential equations\/}.  \publ{Springer Verlag}.

\bibitem[Chan {\em et~al.\/}(2021)Chan, Lee, Pedersen, Dalnoki-Veress \&
  Carlson]{Chan2021}
{\sc \au{Chan, T.S.}, \au{Lee, C.L.}, \au{Pedersen, C.}, \au{Dalnoki-Veress,
  K.} \& \au{Carlson, A.}} \yr{2021}  \at{Film coating by directional droplet
  spreading on fibers}.  \jt{Phys. Rev. Fluids}  \bvol{6}~(1),  \pg{014004}.

\bibitem[Chan {\em et~al.\/}(2020)Chan, Yang \& Carlson]{Chan2020}
{\sc \au{Chan, T.S.}, \au{Yang, F.} \& \au{Carlson, A.}} \yr{2020}
  \at{Directional spreading of a viscous droplet on a conical fibre}.  \jt{J.
  Fluid Mech.}  \bvol{894}.

\bibitem[Chang {\em et~al.\/}(2015)Chang, Bostwick, Daniel \& Steen]{Chang2015}
{\sc \au{Chang, C.-T.}, \au{Bostwick, J.B.}, \au{Daniel, S.} \& \au{Steen,
  P.H.}} \yr{2015}  \at{Dynamics of sessile drops. {P}art 2. {E}xperiment}.
  \jt{J. Fluid Mech.}  \bvol{768},  \pg{442--467}.

\bibitem[Christianto {\em et~al.\/}(2022)Christianto, Rahmawan, Semprebon \&
  Kusumaatmaja]{Christianto2022}
{\sc \au{Christianto, R.}, \au{Rahmawan, Y.}, \au{Semprebon, C.} \&
  \au{Kusumaatmaja, H.}} \yr{2022} Modelling the dynamics of partially wetting
  droplets on fibres,  \arxiv{arXiv: 2209.10329}.

\bibitem[Costalonga \& Brunet(2020)]{Costalonga2020}
{\sc \au{Costalonga, M.} \& \au{Brunet, P.}} \yr{2020}  \at{Directional motion
  of vibrated sessile drops: a quantitative study}.  \jt{Phys. Rev. Fluids}
  \bvol{5}~(2),  \pg{023601}.

\bibitem[Dawar \& Chase(2008)]{Dawar2008}
{\sc \au{Dawar, S.} \& \au{Chase, G.G.}} \yr{2008}  \at{Drag correlation for
  axial motion of drops on fibers}.  \jt{Sep. Purif. Technol.}  \bvol{60}~(1),
  \pg{6--13}.

\bibitem[Dawar {\em et~al.\/}(2006)Dawar, Li, Dobson \& Chase]{Dawar2006}
{\sc \au{Dawar, S.}, \au{Li, H.}, \au{Dobson, J.} \& \au{Chase, G.G.}}
  \yr{2006}  \at{Drag correlation of drop motion on fibers}.  \jt{Dry.
  Technol.}  \bvol{24}~(10),  \pg{1283--1288}.

\bibitem[Deegan(2020)]{Deegan2020}
{\sc \au{Deegan, R.D.}} \yr{2020}  \at{Climbing a slippery slope}.  \jt{J.
  Fluid Mech.}  \bvol{882}.

\bibitem[Ding {\em et~al.\/}(2018)Ding, Zhu, Gao \& Lu]{Ding2018}
{\sc \au{Ding, H.}, \au{Zhu, X.}, \au{Gao, P.} \& \au{Lu, X.-Y.}} \yr{2018}
  \at{Ratchet mechanism of drops climbing a vibrated oblique plate}.  \jt{J.
  Fluid Mech.}  \bvol{835}.

\bibitem[Duprat {\em et~al.\/}(2012)Duprat, Protiere, Beebe \&
  Stone]{Duprat2012}
{\sc \au{Duprat, C.}, \au{Protiere, S.}, \au{Beebe, A.Y.} \& \au{Stone, H.A.}}
  \yr{2012}  \at{Wetting of flexible fibre arrays}.  \jt{Nature}
  \bvol{482}~(7386),  \pg{510--513}.

\bibitem[Gilet {\em et~al.\/}(2009)Gilet, Terwagne \& Vandewalle]{Gilet2009}
{\sc \au{Gilet, T.}, \au{Terwagne, D.} \& \au{Vandewalle, N.}} \yr{2009}
  \at{Digital microfluidics on a wire}.  \jt{Appl. Phys. Lett.}  \bvol{95}~(1),
   \pg{014106}.

\bibitem[Gilet {\em et~al.\/}(2010)Gilet, Terwagne \& Vandewalle]{Gilet2010}
{\sc \au{Gilet, T.}, \au{Terwagne, D.} \& \au{Vandewalle, N.}} \yr{2010}
  \at{Droplets sliding on fibres}.  \jt{Eur. Phys. J. E}  \bvol{31}~(3),
  \pg{253--262}.

\bibitem[Gupta {\em et~al.\/}(2021)Gupta, Konicek, King, Iqtidar, Yeganeh \&
  Stone]{Gupta2021}
{\sc \au{Gupta, A.}, \au{Konicek, A.R.}, \au{King, M.A.}, \au{Iqtidar, A.},
  \au{Yeganeh, M.S.} \& \au{Stone, H.A.}} \yr{2021}  \at{Effect of gravity on
  the shape of a droplet on a fiber: Nearly axisymmetric profiles with
  experimental validation}.  \jt{Phys. Rev. Fluids}  \bvol{6}~(6),
  \pg{063602}.

\bibitem[Ju {\em et~al.\/}(2012)Ju, Bai, Zheng, Zhao, Fang \& Jiang]{Ju2012}
{\sc \au{Ju, J.}, \au{Bai, H.}, \au{Zheng, Y.}, \au{Zhao, Y.}, \au{Fang, R.} \&
  \au{Jiang, L.}} \yr{2012}  \at{A multi-structural and multi-functional
  integrated fog collection system in cactus}.  \jt{Nat. Commmun.}
  \bvol{3}~(1),  \pg{1--6}.

\bibitem[Ju {\em et~al.\/}(2014)Ju, Zheng \& Jiang]{Ju2014}
{\sc \au{Ju, J.}, \au{Zheng, Y.} \& \au{Jiang, L.}} \yr{2014}  \at{Bioinspired
  one-dimensional materials for directional liquid transport}.  \jt{Acc. Chem.
  Res.}  \bvol{47}~(8),  \pg{2342--2352}.

\bibitem[Klemm {\em et~al.\/}(2012)]{Klemm2012}
{\sc \au{Klemm, O.} \& \au{others}} \yr{2012}  \at{Fog as a fresh-water
  resource: overview and perspectives}.  \jt{Ambio}  \bvol{41}~(3),
  \pg{221--234}.

\bibitem[Kovacic {\em et~al.\/}(2018)Kovacic, Rand \& Sah]{Kovacic2018}
{\sc \au{Kovacic, I.}, \au{Rand, R.} \& \au{Sah, S.~M.}} \yr{2018}
  \at{Mathieu's equation and its generalizations: overview of stability charts
  and their features}.  \jt{Appl. Mech. Rev.}  \bvol{70}~(2).

\bibitem[Labb{\'e} \& Duprat(2019)]{Labbe2019}
{\sc \au{Labb{\'e}, R.} \& \au{Duprat, C.}} \yr{2019}  \at{Capturing aerosol
  droplets with fibers}.  \jt{Soft Matter}  \bvol{15}~(35),  \pg{6946--6951}.

\bibitem[Lamb(1924)]{Lamb1924}
{\sc \au{Lamb, H.}} \yr{1924} {\em Hydrodynamics\/}.  \publ{University Press}.

\bibitem[Limm {\em et~al.\/}(2009)Limm, Simonin, Bothman \& Dawson]{Limm2009}
{\sc \au{Limm, E.B.}, \au{Simonin, K.A.}, \au{Bothman, A.G.} \& \au{Dawson,
  T.E.}} \yr{2009}  \at{Foliar water uptake: a common water acquisition
  strategy for plants of the redwood forest}.  \jt{Oecologia}  \bvol{161}~(3),
  \pg{449--459}.

\bibitem[Lorenceau \& Qu{\'e}r{\'e}(2004)]{Lorenceau2004}
{\sc \au{Lorenceau, É.} \& \au{Qu{\'e}r{\'e}, D.}} \yr{2004}  \at{Drops on a
  conical wire}.  \jt{J. Fluid Mech.}  \bvol{510},  \pg{29--45}.

\bibitem[Malik {\em et~al.\/}(2014)Malik, Clement, Gethin, Krawszik \&
  Parker]{Malik2014}
{\sc \au{Malik, F.T.}, \au{Clement, R.M.}, \au{Gethin, D.T.}, \au{Krawszik, W.}
  \& \au{Parker, A.R.}} \yr{2014}  \at{Nature's moisture harvesters: a
  comparative review}.  \jt{Bioinspir. Biomim.}  \bvol{9}~(3),  \pg{031002}.

\bibitem[McCarthy {\em et~al.\/}(2019)McCarthy, Vella \&
  Castrej{\'o}n-Pita]{McCarthy2019}
{\sc \au{McCarthy, J.}, \au{Vella, D.} \& \au{Castrej{\'o}n-Pita, A.A.}}
  \yr{2019}  \at{Dynamics of droplets on cones: self-propulsion due to
  curvature gradients}.  \jt{Soft Matter}  \bvol{15}~(48),  \pg{9997--10004}.

\bibitem[Nayfeh \& Mook(2008)]{NayfehBook}
{\sc \au{Nayfeh, A.H.} \& \au{Mook, D.T.}} \yr{2008} {\em Nonlinear
  oscillations\/}.  \publ{John Wiley \& Sons}.

\bibitem[Noblin {\em et~al.\/}(2009)Noblin, Kofman \& Celestini]{Noblin2009}
{\sc \au{Noblin, X.}, \au{Kofman, R.} \& \au{Celestini, F.}} \yr{2009}
  \at{Ratchetlike motion of a shaken drop}.  \jt{Phys. Rev. Lett.}  \bvol{102},
   \pg{194504}.

\bibitem[Pan {\em et~al.\/}(2016)Pan, Pitt, Zhang, Wu, Tao \&
  Truscott]{Pan2016}
{\sc \au{Pan, Z.}, \au{Pitt, W.G.}, \au{Zhang, Y.}, \au{Wu, N.}, \au{Tao, Y.}
  \& \au{Truscott, T.T.}} \yr{2016}  \at{The upside-down water collection
  system of \emph{{S}yntrichia caninervis}}.  \jt{Nat. Plants}  \bvol{2}~(7),
  \pg{1--5}.

\bibitem[Qu{\'e}r{\'e}(1999)]{Quere1999}
{\sc \au{Qu{\'e}r{\'e}, D.}} \yr{1999}  \at{Fluid coating on a fiber}.
  \jt{Annu. Rev. Fluid Mech.}  \bvol{31}~(1),  \pg{347--384}.

\bibitem[Roth-Nebelsick {\em et~al.\/}(2012)Roth-Nebelsick, Ebner, Miranda,
  Gottschalk, Voigt, Gorb, Stegmaier, Sarsour, Linke \& Konrad]{Roth2012}
{\sc \au{Roth-Nebelsick, A.}, \au{Ebner, M.}, \au{Miranda, T.}, \au{Gottschalk,
  V.}, \au{Voigt, D.}, \au{Gorb, S.}, \au{Stegmaier, T.}, \au{Sarsour, J.},
  \au{Linke, M.} \& \au{Konrad, W.}} \yr{2012}  \at{Leaf surface structures
  enable the endemic namib desert grass \emph{{S}tipagrostis sabulicola} to
  irrigate itself with fog water}.  \jt{J. R. Soc. Interface}  \bvol{9}~(73),
  \pg{1965--1974}.

\bibitem[Sahu {\em et~al.\/}(2013)Sahu, Sinha-Ray, Yarin \&
  Pourdeyhimi]{Sahu2013}
{\sc \au{Sahu, R.P.}, \au{Sinha-Ray, S.}, \au{Yarin, A.L.} \& \au{Pourdeyhimi,
  B.}} \yr{2013}  \at{Blowing drops off a filament}.  \jt{Soft Matter}
  \bvol{9}~(26),  \pg{6053--6071}.

\bibitem[Sartori {\em et~al.\/}(2019)Sartori, Guglielmin, Ferraro, Filippi,
  Zaltron, Pierno \& Mistura]{Sartori2019}
{\sc \au{Sartori, P.}, \au{Guglielmin, E.}, \au{Ferraro, D.}, \au{Filippi, D.},
  \au{Zaltron, A.}, \au{Pierno, M.} \& \au{Mistura, G.}} \yr{2019}  \at{Motion
  of newtonian drops deposited on liquid-impregnated surfaces induced by
  vertical vibrations}.  \jt{J. Fluid Mech.}  \bvol{876}.

\bibitem[Yarin {\em et~al.\/}(2002)Yarin, Liu \& Reneker]{Yarin2002}
{\sc \au{Yarin, A.L.}, \au{Liu, W.} \& \au{Reneker, D.H.}} \yr{2002}
  \at{Motion of droplets along thin fibers with temperature gradient}.  \jt{J.
  Appl. Phys.}  \bvol{91}~(7),  \pg{4751--4760}.

\bibitem[Zhang {\em et~al.\/}(2015)Zhang, Liu, Williams, Qu, Feng \&
  Chen]{Zhang2015}
{\sc \au{Zhang, K.}, \au{Liu, F.}, \au{Williams, A.J.}, \au{Qu, X.}, \au{Feng,
  J.J.} \& \au{Chen, C.-H.}} \yr{2015}  \at{Self-propelled droplet removal from
  hydrophobic fiber-based coalescers}.  \jt{Phys. Rev. Lett.}  \bvol{115}~(7),
  \pg{074502}.

\bibitem[Zhang {\em et~al.\/}(2018)Zhang, Lin \& Yin]{Zhang2018}
{\sc \au{Zhang, Q.}, \au{Lin, G.} \& \au{Yin, J.}} \yr{2018}  \at{Highly
  efficient fog harvesting on superhydrophobic microfibers through droplet
  oscillation and sweeping}.  \jt{Soft Matter}  \bvol{14}~(41),
  \pg{8276--8283}.

\bibitem[Zheng {\em et~al.\/}(2010)Zheng, Bai, Huang, Tian, Nie, Zhao, Zhai \&
  Jiang]{Zheng2010}
{\sc \au{Zheng, Y.}, \au{Bai, H.}, \au{Huang, Z.}, \au{Tian, X.}, \au{Nie,
  F.-Q.}, \au{Zhao, Y.}, \au{Zhai, J.} \& \au{Jiang, L.}} \yr{2010}
  \at{Directional water collection on wetted spider silk}.  \jt{Nature}
  \bvol{463}~(7281),  \pg{640--643}.

\end{thebibliography}

\end{document}